\documentclass[fleqn,usenatbib,useAMS]{mnras}

\usepackage{graphicx}	
\usepackage{float}
\usepackage{seqsplit}
\usepackage{amsmath}	
\usepackage{amssymb}	
\usepackage{multicol}        
\usepackage{bm}		
\usepackage{pdflscape}	

\usepackage{color}

\newcommand{\cm}{\ensuremath{\,{\rm cm}}}
\newcommand{\m}{\ensuremath{\,{\rm m}}}
\newcommand{\km}{\ensuremath{\,{\rm km}}}
\newcommand{\kms}{\ensuremath{\,\km}\,s^{-1}} 
\newcommand{\Mpc}{\ensuremath{\,{\rm Mpc}}}

\newcommand{\GHz}{\ensuremath{\, {\rm GHz}}}
\newcommand{\Jy}{\ensuremath{\,{\rm Jy}}}

\newcommand{\Msun}{\ensuremath{\, {\rm M}_{\odot}}}
\renewcommand{\deg}{\ensuremath{\,{\rm deg}}}
\renewcommand{\arcmin}{\ensuremath{\,{\rm arcmin}}}

\usepackage{array}
\newcommand{\PreserveBackslash}[1]{\let\temp=\\#1\let\\=\temp}
\newcolumntype{C}[1]{>{\PreserveBackslash\centering}p{#1}}
\newcolumntype{R}[1]{>{\PreserveBackslash\raggedleft}p{#1}}
\newcolumntype{L}[1]{>{\PreserveBackslash\raggedright}p{#1}}

\usepackage[T1]{fontenc}
\usepackage{ae,aecompl}

\usepackage{newtxtext,newtxmath}

\title{\textit{Forecast for FAST: from Galaxies Survey to Intensity Mapping}}

\author[Wenkai Hu et al.]
{Wenkai Hu$^{1,2}$,
Xin Wang$^3$, 
Fengquan Wu$^{1}$, 
Yougang Wang$^{1}$, 
Pengjie Zhang$^{4}$,
\newauthor
Xuelei Chen$^{1,2,5}$\thanks{Contact e-mail: \href{xuelei@cosmology.bao.ac.cn}{xuelei@cosmology.bao.ac.cn}}
\\
$^{1}$ Key Laboratory of National Astronomical Observatories, Chinese Academy of Sciences, Beijing 100012, China\\
$^{2}$ School of Astronomy and Space Science, University of Chinese Academy of Sciences, Beijing 100049, China\\
$^{3}$ School of Physics and Astronomy, Sun Yat-Sen University, Guangzhou 510297, China\\
$^{4}$ Department of Astronomy, School of Physics and Astronomy, Shanghai Jiao Tong University, Shanghai, 200240, China\\
$^{5}$ Center of High Energy Physics, Peking University, Beijing 100871, China
}
\date{Last updated September 24, 2019}

\pubyear{2019}

\begin{document}
\label{firstpage}
\pagerange{\pageref{firstpage}--\pageref{lastpage}}
\maketitle

\begin{abstract}
The Five-Hundred-Meter Aperture Spherical Radio Telescope(FAST) is the largest single-dish radio telescope in the world. In this paper, 
we make forecast on the FAST HI large scale structure survey by mock observations. We consider a drift scan survey with the L-band 19 beam 
receiver, which may be commensal with the pulsar search and Galactic HI survey. We also consider surveys at lower frequency, either using 
the current single feed wide band receiver, or a future multi-beam phased array feed (PAF) in the UHF band. We estimate the number density 
of detected HI galaxies and the measurement error in positions, the precision of the surveys are evaluated using both Fisher matrix and 
simulated observations. The measurement error in the HI galaxy power spectrum is estimated, and we find that 
the error is relatively large even at moderate redshifts, as the number of positively detected galaxies drops drastically with increasing redshift. 
However, good cosmological measurement could be obtained with the intensity mapping technique where the large scale HI distribution is 
measured without resolving individual galaxies. The figure of merit (FoM) for the dark energy equation of state with different observation times 
are estimated, we find that with the existing L-band multi-beam receiver, a good measurement of low redshift large scale structure can be obtained, 
which complements the existing optical surveys. With a PAF in the UHF band, the constraint can be much stronger, reaching the level of a
dark energy task force (DETF) stage IV experiment.
\end{abstract}

\begin{keywords}
galaxies: evolution - galaxies: ISM - radio lines: galaxies
\end{keywords}



\section{Introduction}
Over the past two decades, an increasing number of optical galaxy surveys, such as the 2DFGRS \citep{2001MNRAS.327.1297P}, SDSS\citep{1995wfsd.conf....3G,2004ApJ...606..702T,2009astro2010S.314S,2016AJ....151...44D}, WiggleZ\citep{2007ASPC..379...72G,2010MNRAS.401.1429D} are probing increasingly large volumes of the Universe, and provide large scale structure (LSS) data for cosmological studies. For example, assuming that the observed number density of galaxies traces the total density of the 
matter distribution, the baryon acoustic oscillation (BAO) features in the galaxy power spectrum are measured, and used as standard rulers to constrain cosmological models \citep{2005ApJ...633..560E,2005MNRAS.362..505C,2010MNRAS.401.2148P,2011MNRAS.416.3017B,2011MNRAS.418.1707B,2015MNRAS.449..835R,2017MNRAS.470.2617A,2018MNRAS.473.4773A}. While it is plausible that the galaxies formed from over-density perturbations and therefore trace the total density on large scales, it is vital to check this hypothesis and also understand the range of validity of it by observing the galaxies with different means. The 21cm line of the neutral hydrogen (HI) provides a good alternative way of observation in the radio wavelength. A number of HI galaxies surveys have also been carried out, e.g. the HIPASS survey \citep{2004MNRAS.350.1195M,2004MNRAS.350.1210Z}, and the ALFALFA survey \citep{2005AJ....130.2598G,2007AJ....133.2087S,2007AJ....133.2569G}, and JVLA deep survey \citet{2014arXiv1401.4018J}. However, limited by the  sensitivity of the telescopes, the redshift range of these surveys are much smaller than the current optical surveys.

A new generation of radio telescopes are being built or under development, including the Square Kilometer Array (SKA)  
in the southern hemisphere, and the FAST\citep{2011IJMPD..20..989N} in the 
northern hemisphere. These radio telescopes have much better sensitivities and observe HI galaxies at larger 
distances.  Here we consider FAST, which is about the complete its commissioning process and starts science runs.
 It has a very large aperture (300 meters during operation) and is to be equipped with multi-beam feed system and low-noise cryogenic
 receivers, ideal for conducting large surveys. 

The FAST has unprecedented large effective area and high sensitivity, nevertheless for a traditional 
galaxy survey \citep{2008MNRAS.383..150D} the redshift at which an individual galaxy could be detected is still very
limited, and its angular resolution would be insufficient to resolve the galaxy at high redshift. However,
to map the large-scale structure, in principle it is not necessary to resolve individual galaxies as 
traditional galaxy surveys do, instead the redshifted 21cm line intensity can 
be mapped with lower angular resolution, as is done in the Epoch of Reionization (EoR) 
experiments\footnote{One of us (XC) first realized that this mode of observation could be used with FAST to probe the large scale
structure and presented it at a meeting on the FAST science case held in May 2007 in Hangzhou, China.}.
In more general context, \citet{2008PhRvL.100i1303C} studied this mode of observation and named it the {\it intensity 
mapping} method, and  also proposed that  a cost-effective way to survey large scale structure is to 
develop a dedicated dense array of cylinder or small dish 
antennas \citep{2008PhRvL.100i1303C,2008arXiv0807.3614A,2010ApJ...721..164S,2012A&A...540A.129A}. Indeed
a number of such small-to-mid scale experiments are undergoing, such as those of 
Tianlai \citep{2012IJMPS..12..256C,2015ApJ...798...40X}, CHIME \citep{2014SPIE.9145E..22B} and 
HIRAX\citep{2016SPIE.9906E..5XN}, as well as the specially designed single dish experiment 
BINGO \citep{2012arXiv1209.1041B,2016arXiv161006826B}. 

For the FAST itself, several studies used the Fisher matrix formalism to make simple forecasts on the constraining power of 
cosmological parameters by HI galaxy survey \citep{2008MNRAS.383..150D} or intensity mapping surveys 
\citep{2015MNRAS.454.3240B,2017A&A...597A.136S,2019arXiv190803024Y}.

In this paper, we make a more detailed investigation by 
simulating the observed galaxies, and also compare the galaxy survey and intensity mapping. 
The layout of this paper is as follows. In Sec.II we describe our model of the telescope and its receiver feeds. the 
In section III, we present the modeling of the HI galaxies and their observation, as well as the simulated intensity map.  In section IV,  we make Fisher matrix
forecasts of the precision of power spectrum measurement using both HI galaxy surveys and intensity mapping surveys, and also make measurement using 
numerical simulation. The niche of HI galaxy survey and HI intensity mapping survey and the effect of foreground are discussed in Sec. V. Finally we summarize
the results in section VI.

\section{The FAST telescope}
In order to study how the FAST could survey the large-scale-structure, we conduct mock observations with 
simulated sky. We first generate a catalog of galaxies from simulation, then convert it into the simulated 
sky of HI intensity as would be observed by FAST. 

\subsection{The Instrument\label{sec:instr}}

The diameter of the FAST reflector is 500 m, the fully illuminated aperture at any time is $D=300$ m
since the telescope is designed to track 
objects. The beam size of the FAST is given by
\begin{eqnarray}
\theta &=& 1.22 \times \frac{21\cm (1+z)}{300 \m}= 2.94 (1+z) \arcmin	\end{eqnarray}
for observation of the 21cm line from redshift $z$.
 
During a drift scan, a single feed is fixed to be pointed to a particular declination in the due north or south direction, 
so that in a sidereal day, a ring of width $2.94(1+z) \arcmin$ centered at that declination is scanned. The pointing declination 
can be changed so as to cover the whole observable part of the sky. The FAST site is located at a latitude of 25$^{\circ}$48$^{\prime}$ North, 
and the maximum zenith angle is $40^\circ$,  allowing the observation of $\approx$ 50$\%$ of the full sky or about 20,000 $\deg^2$. 
  
 The FAST is equipped with a number of different feed and receiver systems. For HI survey, the most relevant 
are the L-band 19-beam feed/receiver system and 
the wide band receiver system. Additionally, there are also several low frequency receivers which cover down to 70 MHz,
which can be used for Epoch of Reionization (EoR) observations. Here we shall consider mainly the first two, which are relevant for 
low-or-mid redshift observations of large scale structure.  In addition, below we shall also consider a possible future UHF phased 
array feed (PAF) system. We summarize the information of these receiver systems in Table~\ref{receiver_table}. 

{\bf L-band 19 beam receiver system}. 
It covers the frequency range of 1.05-1.45 GHz, and the beams are arranged in two 
concentric hexagonal rings around the central beam. 
The minimum spacing between beam centers is 5.73 arcmin and is approximately constant, though for each beam the width scales roughly as $\theta \propto (1+z)$.
In this paper we assume the feed array are tilted an angle of 23.4 degree
with respect to the compass points to increase the area covered for each scan, 
as was proposed for the Commensal Radio Astronomy FasT survey (CRAFTS)\citep{2018IMMag..19..112L}, though we note that this is not the only choice available.
The whole 19 beams span 22.8 arcmins across the north-south direction at 1.42GHz (calculated for the centre of the beam).  The sky is covered by shifting the whole array in declination by 21.9 arcmin for the next scan.
A drift scan of $\pm 40^\circ$ from the center declination would require about 220 strips 
(i.e. 220 days)  to cover the region once. 



\begin{table}
	\centering
	\caption{FAST survey receiver parameters. The $t_{\rm sur}$ refers to the time needed to finish a full drift scan of $\pm 40^\circ$ from 
	the center declination of FAST. The $T_{\rm rec}$ is the receiver noise.}
	\label{receiver_table}
	\begin{tabular}{ccccc} 
		\hline
		receiver & band(GHz) &  Beams &$T_{\rm rec}$(K)& $t_{\rm sur}$(days)\\
		\hline
		L-band & 1.05-1.45 & 19 & 20 & 220\\
        Wide-band & 0.27-1.62 & 1 & 60 & 1211\\
        UHF PAF (future) & 0.5-1.0 & 81 & 30 & 135 \\
		\hline
	\end{tabular}
\end{table}

{\bf Wide-band receiver system.} 
For higher redshift (z>0.35), at present the survey can be done with a single feed wide band receiver, which covers a frequency 
from 0.27GHz to 1.62 GHz.The receiver noise for Wide-band receiver system is $\approx$ 60K, to have the same noise scale, the survey for redshift larger than 0.35 need twice more time than the survey for redshift smaller than 0.35.  The strip width in this case is $2.9(1+z)\arcmin$, so to 
cover the $\pm 40 \deg$ sky, it would require 1211 days with single feed strip to cover the same sky region at $z=0.35$, which is much less practical due to the 
long observation time required. 

{\bf The PAF receiver system for UHF band.}
 In the future, it is worthwhile to consider equipping the FAST telescope 
with a multi-beam receiver at the lower frequency band for a survey of higher redshifts. 
A phased array feeds (PAF) with cryogenic receiver system would allow rapid survey of large areas of sky, and such 
development has also been pursued for FAST  \citep{wu2016}. 
Here as an illustrative example, we consider a low frequency PAF system with 81 effective beams, 500MHz bandwidth centered at 0.75 GHz,  
(i.e. 0.5-1.0 GHz), a system noise about 30K,  and an aperture efficiency around 70$\%$. These beams could positioned in square array, then a pixel in the sky 
will be scanned 9 times if drift along a side of the square. In this way, a full drift scan of $\pm$40$^{\circ}$ could be finished in 135 days, 
with integration time of 291s on each pixel.

\subsection{Integration Time and Noise}
\begin{figure}
  \centering
     \includegraphics[width=0.45\textwidth]{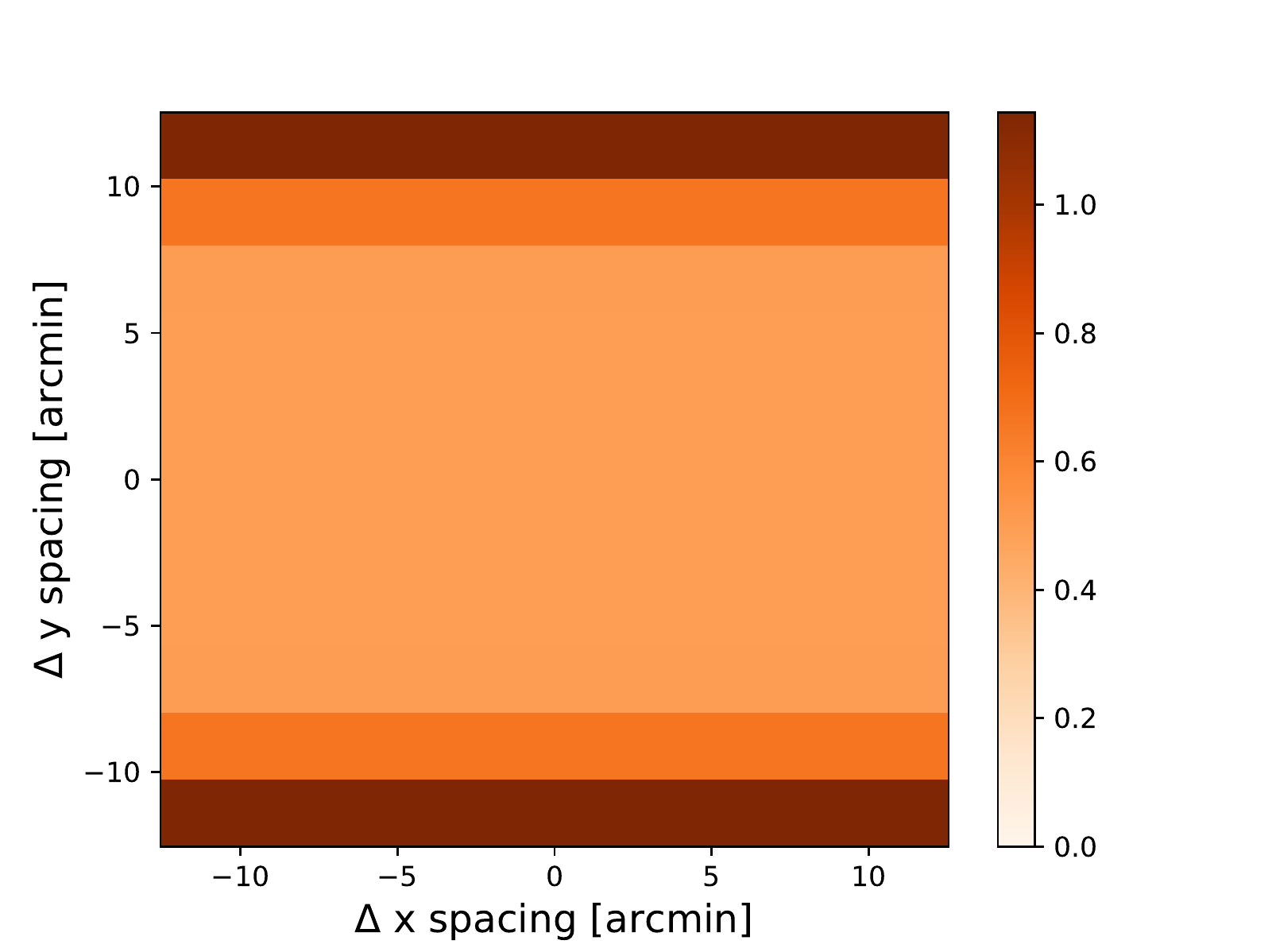}
     \includegraphics[width=0.45\textwidth]{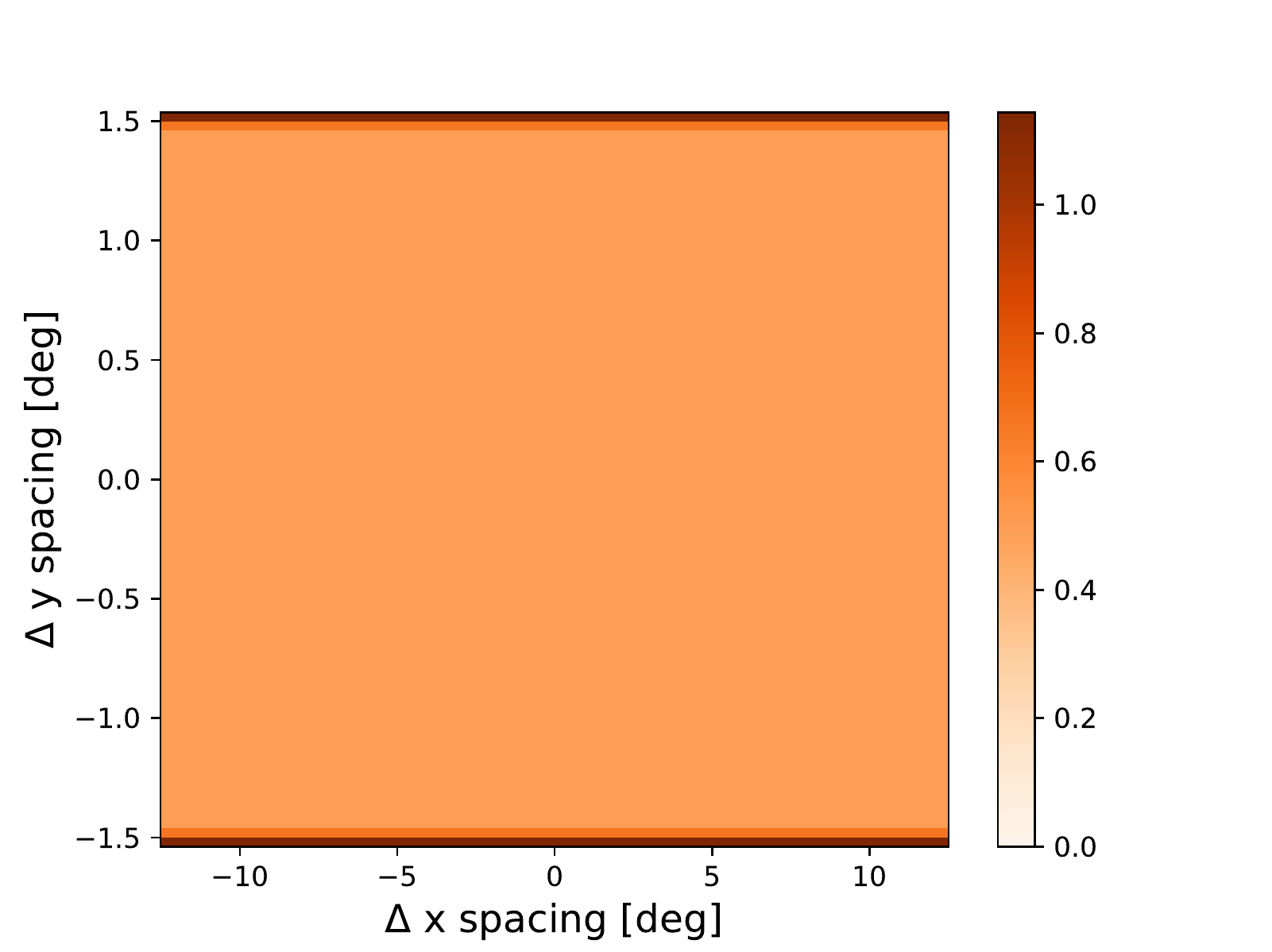}
  \caption{The top panel shows the strips of noise level produced by one scan of the 19-beam feed. The bottom panel shows the noise level from multiple scans with vertical intervals of 21.9 arcmin. The width of strip is selected to be 2.3 arcmin, which is the vertical intervals of two nearest dashed line in the top panel.}
  \label{noise_strip}
\end{figure}

The sky drifts across with a speed of $\omega_e \cos \delta$ in a drift scan survey,  where 
$\omega_e \approx 0.25 \arcmin/s$ is the angular velocity of the rotation of the
Earth, and $\delta$ is the declination of the pointing.  The time for drifting across a pixel is given by 
\begin{eqnarray}
  t_{\rm pix}&=& 2.9(1+z) \arcmin /(\omega_e \cos\delta).
\end{eqnarray}
One circle is completed in a sidereal day, though in practice the night time data is usually of 
much smaller noise than the day time data.  
At $z=0$, and $\cos\delta \approx 1/2$ (near the zenith of FAST site), we get 24s per beam. And because of the overlap of 19 beams in one horizontal scanning(see Fig. \ref{noise_strip}), most pixels in a 19-beams strip will be scanned twice, resulting in 48s per beam. Within the observable part of sky, the circles with 
higher declination (northern part of sky) has smaller area, while the integration time per pixel is larger. The expected thermal noise for a dual polarization single beam is
\begin{eqnarray}
\sigma_{\rm noise} = \sqrt{2}\frac{k_B T_{\rm sys}}{A_{\rm eff}}\frac{1}{\sqrt{\Delta\nu t}}
\label{thermal_noise}
\end{eqnarray}
where $t$ is the total integration time, $\Delta\nu$ is the frequency bandwidth for a channel, and $k_B$ the Boltzmann constant. The aperture efficiency is about 70$\%$, giving effective aperture $A_{\rm eff} \approx$ 50000 $\m^2$. 
The system temperature is $T_{\rm sys}= T_{\rm rec} + T_{\rm sky}$, 
where $T_{\rm rec}$ is the receiver temperature and taken to be 20K, and away from the Galactic plane, the sky temperature is modeled as 
\begin{eqnarray}
T_{\rm sky} = 2.73 + 25.2 \times(0.408/\nu_{\rm GHz})^{2.75} K.
\label{sky_noise}
\end{eqnarray}

 If we assume a velocity line width of $\approx$ 5 $\kms$ for the 
spectral line observation in HI galaxy, and 48s integration time per beam, the instantaneous sensitivity of each beam of the FAST system will be 0.86mJy. Below we shall consider surveys with an average of 48s, 96, 192s and 384s integration time per beam, according to once, twice, three times and  four times repeat observation respectively.

In the case of 19 beam L-band feed,  approximately every pixel would be covered by several beams, effectively double the integration time. In a more careful treatment,  we may estimate the noise as follows. The time stream data is related to the signal by 
\begin{eqnarray}
\mathbf{d={A s} + n},
\end{eqnarray}
where the time-ordered data vector $\mathbf{d}$ has a dimension of $19 N_t$ 
where $N_t$ is the length of the time-ordered-data, the sky pixel vector $\mathbf{s}$ has a dimension $N_{\rm pix}$, 
and the pointing matrix $\mathbf{A}$ has a dimension of $19 N_t \times N_{\rm pix}$. 
The minimum variance estimator for the sky is 
\begin{eqnarray}
\mathbf{ \hat{s} = (A^t N^{-1} A)^{-1} A^{t} N^{-1} s}
\end{eqnarray}
where $\mathbf{N}$ is the covariance matrix of the noise in the time-ordered data.
The sky map noise covariance matrix is then 
\begin{eqnarray}
    \mathbf{C_{N} = (A^{T}N^{-1}A)^{-1}}.
\label{noise_matrix}
\end{eqnarray}
Using this expression we can estimate the map noise.  

In Fig. \ref{noise_strip}, we show the estimated noise of the sky map obtained by the 19-beam receiver in units of 
single beam receiver. For simplicity we assumed the beams are identical and have a gaussian beam within the beam width, 
though in reality there is much difference in the central and outer beams. Also, we assumed a constant system temperature, though actually the system temperature varies, as the sky temperature varies. 

As one might expect, in a scan along the horizontal direction of the 
19 beam receiver, pixels which are near the center of the receivers will be scanned by more than one beams, 
resulting in a lower noise than others. Such inhomogeneous noise distribution is undesirable, because in the large scale structure
measurement it may bias the observation and induce superfluous structures. The bias may be approximately corrected by
introducing selection functions, but as the real noise is varying and not accurately known, precision is hard to achieve. To reduce 
such effects, we need to have a relatively uniform distribution of noise in the survey regions. From Fig. \ref{noise_strip},
we see by partially overlap the scanning strips (with a vertical intervals of 21.9 arcmin), a large part of this inhomogeneity 
could be removed, making it a nearly uniform survey in the central part.
\section{Simulation}

\subsection{The Galaxy Model}
\begin{figure}
  \centering
  \includegraphics[width=0.45\textwidth]{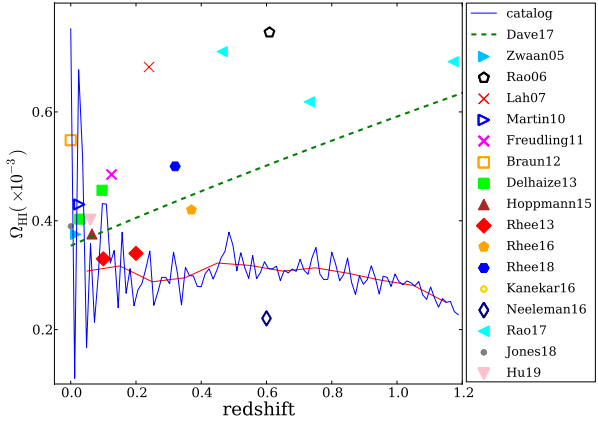}
  \caption{The $\Omega_{\rm HI}(z)$ from the mock catalog (blue line, and red line for the smoothed result), and the 
  various observations in . The dashed green line represents results in 
  the MUFASA cosmological hydrodynamical simulation from: $\Omega_{\rm HI}= 10^{-3.45}\times(1+z)^{0.74}$.}
  \label{omega_fit}
\end{figure}

We used the catalog from the Semi-Analytic Suite of the SKA Simulated Skies($S^3$-SAX), in which 
the cosmic evolution of the galaxies is tracked by semi-analytic models \citep{2007MNRAS.375....2D} based on the 
Millennium N-body simulation \citep{2005Natur.435..629S}, and the amount of neutral atomic hydrogen (HI)
and molecular (H$_{2}$) hydrogen in galaxies are computed with the semi-analytical 
model \citep{2009ApJ...698.1467O,2009ApJ...702.1321O,2009ApJ...703.1890O}. 
An easy-to-use mock catalog (\citet{2014arXiv1406.0966O}) of galaxies with detailed physical 
properties (position in the sky, apparent redshift, stellar mass, HI mass, effective radius, etc.) is available. The catalog is for a cone 
with a field of 10-by-10 degrees and a redshift range of 0.0-1.2. 
It is complete down to an HI mass of $10^8 \Msun$. A deficiency of this model is that this HI cutoff mass is still relatively high, which could miss a significant amount of HI in dwarf galaxies. 
This limit is only a minor concern when dealing with isolated direct HI detections in blind 
surveys, because only a tiny fraction of the total survey volume is sensitive to HI masses < $10^8 M_{\sun}$. 
However, when dealing with global HI mass estimates, e.g. the intensity mapping experiment, the 
HI mass contained in unresolved galaxies is non-negligible. 
We compute $\Omega_{\rm HI}(z)$ from all the galaxies in the mock catalogue, and compare it with the 
observations \citep{2005MNRAS.359L..30Z,2007MNRAS.376.1357L,2010ApJ...723.1359M,2011ApJ...727...40F,2012ApJ...749...87B,
2013MNRAS.433.1398D,2006ApJ...636..610R,2017MNRAS.471.3428R,2013MNRAS.435.2693R,2016MNRAS.460.2675R,
2018MNRAS.473.1879R,2015MNRAS.452.3726H,2016ApJ...818L..28K,2016ApJ...818..113N,2018MNRAS.tmp..502J,2019MNRAS.489.1619H},  
the result is shown in  Fig. \ref{omega_fit}. There are still quite large scatters and discrepancy in the result, but already we can see
the $S^3$-SAX simulation may have under-estimated the amount of HI by a factor of between 1.2 to 2.0,
especially at higher redshifts. The more recent MUFASA cosmological hydrodynamical simulation \citep{2017MNRAS.467..115D}
is in better agreement 
with the observations. We have done most of our galaxy survey simulation with the $S^3$-SAX mock catalogue, for 
the computation of the intensity mapping, we scale all HI flux(equivalent to HI mass) with a 
z-dependent factor to compensate the lost HI mass.

For each galaxy in the catalog, the HI distribution and 21cm emission is modeled following \citep{2016MNRAS.460.4366E}. 
We generate a mini data cube for each galaxy, then we re-grid the mini data cube to a full-size cube which contains 
all the galaxies . 

The HI mass density distribution of the galaxy is modeled as a thin axisymmetric exponential model:
\begin{eqnarray} 
\Sigma_{\rm HI}(r) = \frac{\tilde{\Sigma}_{\rm H} e^{-r/r_{\rm disk}}}{1+R^{c}_{\rm mol} e^{-1.6r/r_{\rm disk}}},
\label{sigma_HI}
\end{eqnarray}
where $r$ denotes the galactocentric radius, $r_{\rm disk}$ refers to the scale length, $\Sigma_{\rm HI}(r)$ is the surface 
density of the total hydrogen component, $\Sigma_{\rm H} = M_{\rm H} /(2\pi r^{2}_{\rm disk} )$ is a normalization factor 
and $R^{c}_{\rm mol}$ denotes the ${\rm H}_2$ /HI-mass ratio at the galaxy center. 
This is based on a list of empirically supported assumptions, (i) the cold gas of regular galaxies resides in a flat disk 
(see \cite{2008AJ....136.2782L} for local spiral galaxies, \cite{2002AJ....124..788Y} for local elliptical galaxies, 
\cite{2006ApJ...640..228T} for galaxies at higher redshifts); (ii) the surface density of the total hydrogen component (HI+H2) 
is well described by an axisymmetric exponential profile (\cite{2008AJ....136.2782L}); (iii) the local H2/HI-mass ratio scales 
as a power of the gas pressure of the ISM outside molecular clouds (\cite{2006ApJ...650..933B}). Recent 
observation(\cite{2012MNRAS.422.1835S,2017MNRAS.464.1029N}) for early type 
galaxies reveals that for most of the elliptical galaxies, the HI gas has a morphology that is similar in appearance to the 
discs of radio emission associated with SF in spiral galaxies. Most of the HI detection exhibits a large, settled HI disc or ring.
The orientation of the galaxy is randomly chosen. In reality, the galaxy may have some correlation of 
intrinsic alignment in their orientation, but such alignment is generally a second order effect, and does not significantly 
affect the analysis given below.
When generating a galaxy model with the parameters from the catalog, we convert the apparent HI half-mass radius, 
$R_{\rm HI}^{\rm half}$, along the major axis into an exponential disc scale length $r_{\rm disk}$. The galaxy is modeled 
out to a radius of 3.5 $r_{\rm disk}$. With the surface density deduced, we then convert it to the mass distribution. 
The circular velocity profile of the galaxy is modeled with the Polyex analytic function \citep{2002ApJ...571L.107G}: 
\begin{eqnarray}
V_{\rm PE}(r) = V_{0}(1-e^{-r/r_{\rm PE}})(1+\ \frac{\alpha r}{r_{\rm PE}})
\label{circular}
\end{eqnarray}
where $V_{0}$, $r_{\rm PE}$, and $\alpha$  determine the amplitude, exponential scale of the inner region, 
and the slope of the outer part of the rotation curve respectively. These parameters are derived from the luminosity of the 
galaxy given in the semi-analytical model,  using the empirical relations derived from nearly 2200 low redshift disk galaxies\citep{2006ApJ...640..751C}.
The semi-analytical model gives R-band luminosity, while the \citet{2006ApJ...640..751C} model used I-band luminosity, 
so we convert them by $M_{I} = M_{R} - 0.37$ \citep{2012MNRAS.426.3385D}.

The HI flux density is then given by
\begin{eqnarray}
\frac{M_{\rm HI}}{M_{\sun}}= 2.36 \times 10^5 (\frac{D_L}{\Mpc})^2 \frac{S_{i}}{\Jy} \frac{d v}{\kms} (1 + z)^{-2},
\label{HI_flux}
\end{eqnarray}
where $S_{i}$ is flux density in units of Jy in channel i of the mini-cube, dv is the velocity width of a channel in$\kms$, and $D_{L}$ is the luminosity distance of the target galaxy in $\Mpc$ units, and z is its evaluated redshift. Note that $dv$ here is defined in intrinsic velocity bin, 
if it is for the observed velocity bin, the $(1 + z)^{-2}$ will be replaced by $(1 + z)^{-1}$. 
For each galaxy, the  data cube has $100 \times 100$ angular pixels, and $5\kms$ in velocity channel width. Each voxel (volume pixel) have its HI flux density and velocity that computed, and we reposition it in a 3D data cube.


We use a light cone catalogue from the \citet{2014arXiv1406.0966O} simulation, which spans a field of $10\times10 \deg^2$ on the 
sky and a redshift range of 0.0-1.2. This volume contains 19,210,309 galaxies with a total HI mass 
of $2.065 \times 10^{16} M_{\sun}$. 
We then re-grid the mini data cubes into the full-size synthetic cube and place it in the corresponding angular position and frequency. The 
final full-size synthetic cube have a pixel width of 0.0133 deg 
and a fixed channel width of 
0.0237MHz, corresponding to 5$\kms$ at the redshift of z = 0.
To simulate the sky observed by FAST, we convolve each channel of the synthetic data cube with a redshift-dependent Gaussian 
point-spread-function (PSF), with beam width proportional to $1+z$. 


\subsection{HI Galaxy Detection}
To simulate the detection of HI galaxies, we first re-bin the full-size synthetic data in RA and DEC to 
an angular resolution of 0.08degree, corresponding to 2 times FWHM of the FAST beam. 
If we assume the noise for each beam is gaussian, the 
noise for each pixel is rescaled as 
\begin{eqnarray}
    \sigma_{\rm noise}^{\rm pixel} = \sigma_{\rm noise}^{\rm beam}\times\sqrt{\frac{A_{\rm pixel}}{A_{\rm beam}}},
\label{thermal_noise_pixel}
\end{eqnarray}
where $A_{\rm pixel}$ is the sky area of a pixel and $A_{\rm beam}$ is the sky area of the beam.

In the frequency axis, the synthetic data cube is re-binned to a resolution of 0.1MHz, 
corresponding to a velocity width of 20km/s at z = 0. Because the flux and noise scale will change with 
the bandwidth, if the full-size synthetic data is smoothed to a velocity width of $W_{s}$$\kms$, 
the signal-to-noise from one velocity bin is scaled as (\citet{2007AJ....133.2087S})
\begin{eqnarray}
    S/N = \frac{(F_{0}/N)/W_{s}}{\sigma_{0}}\times\bigg(\frac{W_{s}}{5\km \mathrm{s}^{-1}}\bigg)^{1/2},
\label{signal_to_noise}
\end{eqnarray}
where $F_{0}$  is the total velocity integrated HI flux of a galaxy, $N$ is the number of the velocity bins the galaxy spans 
and $\sigma_{0}$ is the thermal noise with a velocity width of 5$\kms$. 

With the above mock data, we may simulate galaxy detection as follows:
\begin{enumerate}
 \item {\it coarse resolution search.} Re-bin the noise filled data to a angular resolution of 0.08$\deg$(two times of FWHM of FAST) 
 and a frequency resolution of 0.473MHz (corresponding to velocity resolution of 100km/s at redshift 0), 
 setting a threshold of $3\sigma$, and detect voxels above the threshold.
 \item {\it fine resolution fit.} For galaxies detected in the coarse search, use a finer frequency resolution (0.0236MHz, corresponding to 
 velocity resolution of 5km/s at redshift 0) to fit its spectrum in the data cube with a parameterized profile function. 
 If a reasonable HI profile is obtained,  we integrate the HI profile, the candidate is selected as a galaxy if the total flux exceeds 5 $\sigma$.
 About  20$\%$ of the candidates found in the first step passed the second step, the other ones might be large noise.
 \end{enumerate} 

%

For the HI profile, we use the so-called "busy function" proposed by  \citet{2014MNRAS.438.1176W}, 
which has great flexibility in 
fitting a wide range of HI profiles from the Gaussian profiles of dwarf galaxies to the broad, asymmetric double-horn profiles of 
spiral galaxies. 

\subsection{Galaxy Distribution}
To show the capacity of surveys with different integration time, we have run our selection pipeline with three different $\sigma_{0}$: 0.86mJy, 0.61mJy, 0.43mJy and 0.31mJy, corresponding to 48s, 96s, 192s and 384s integration time per beam.

 \begin{figure}
  \centering
  \includegraphics[width=0.45\textwidth]{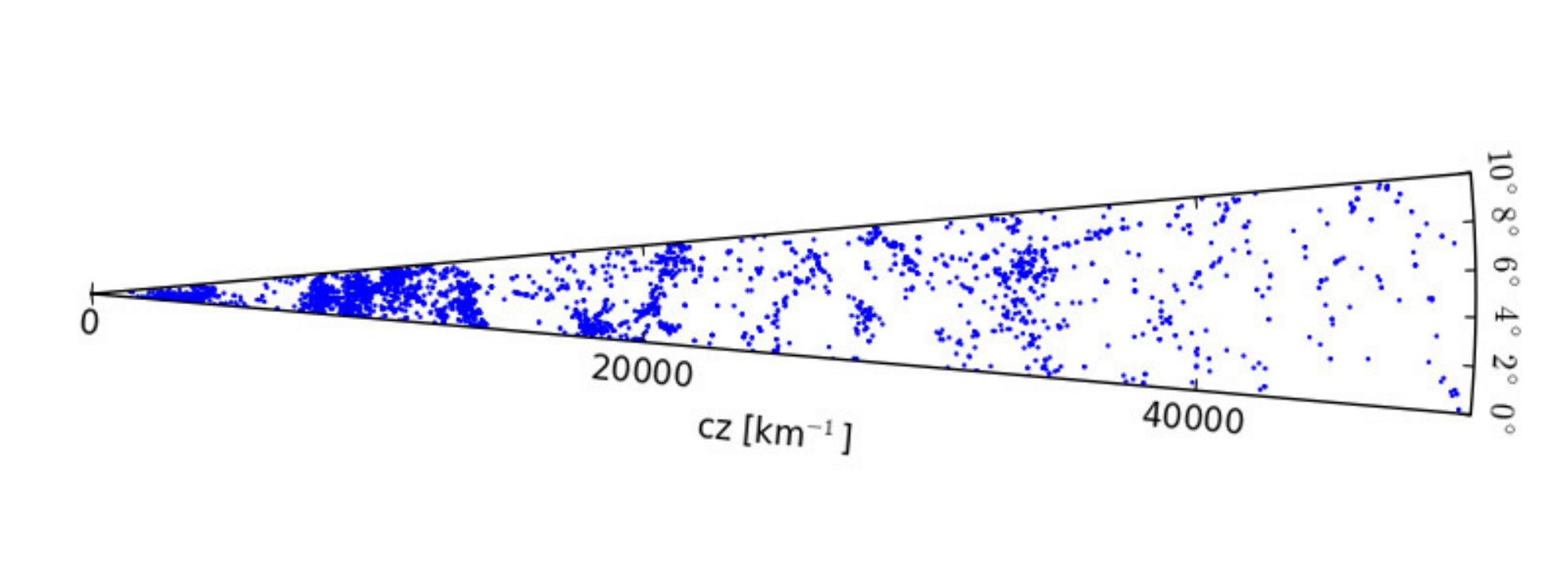}\\
  \includegraphics[width=0.45\textwidth]{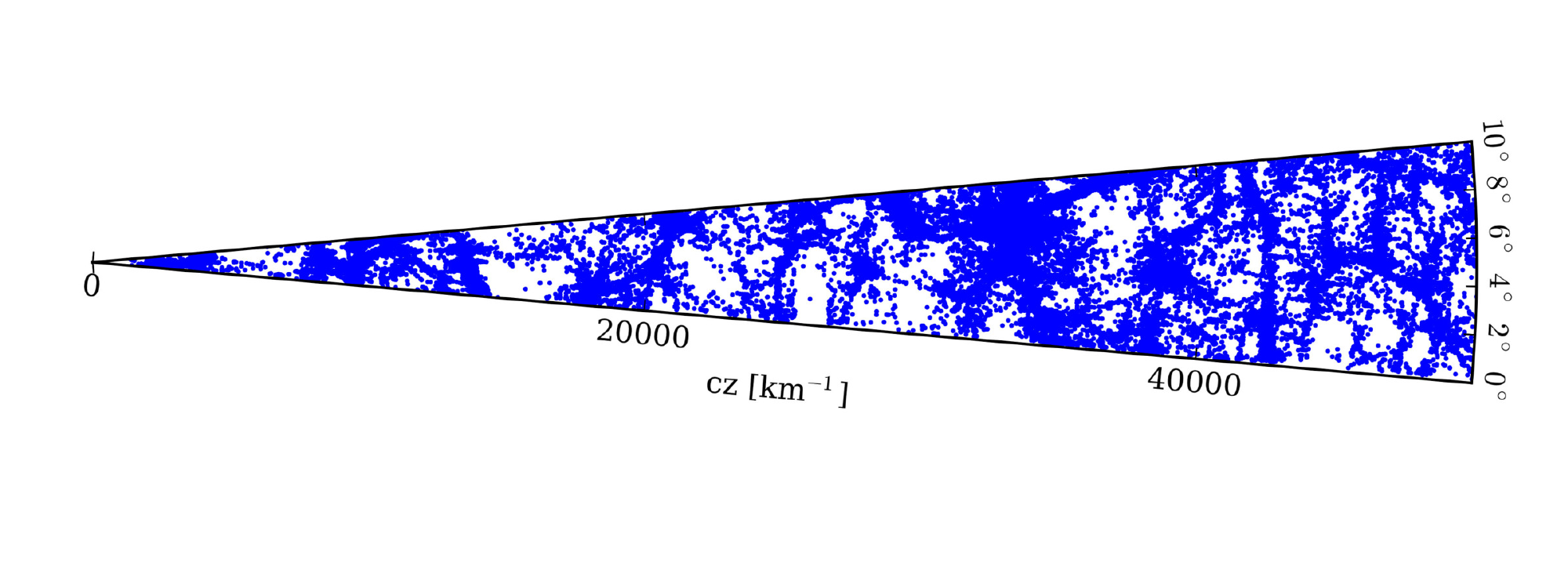}
  \caption{The top panel is the 10 $\Mpc/h$ slab of the galaxy candidates from the FAST 2-years(or 192s-integration-time) galaxy survey, 
  the bottom panel is the distribution of all galaxies in the catalog.}
  \label{polar_distribution}
\end{figure}

\begin{figure}
  \centering
  \includegraphics[width=0.45\textwidth]{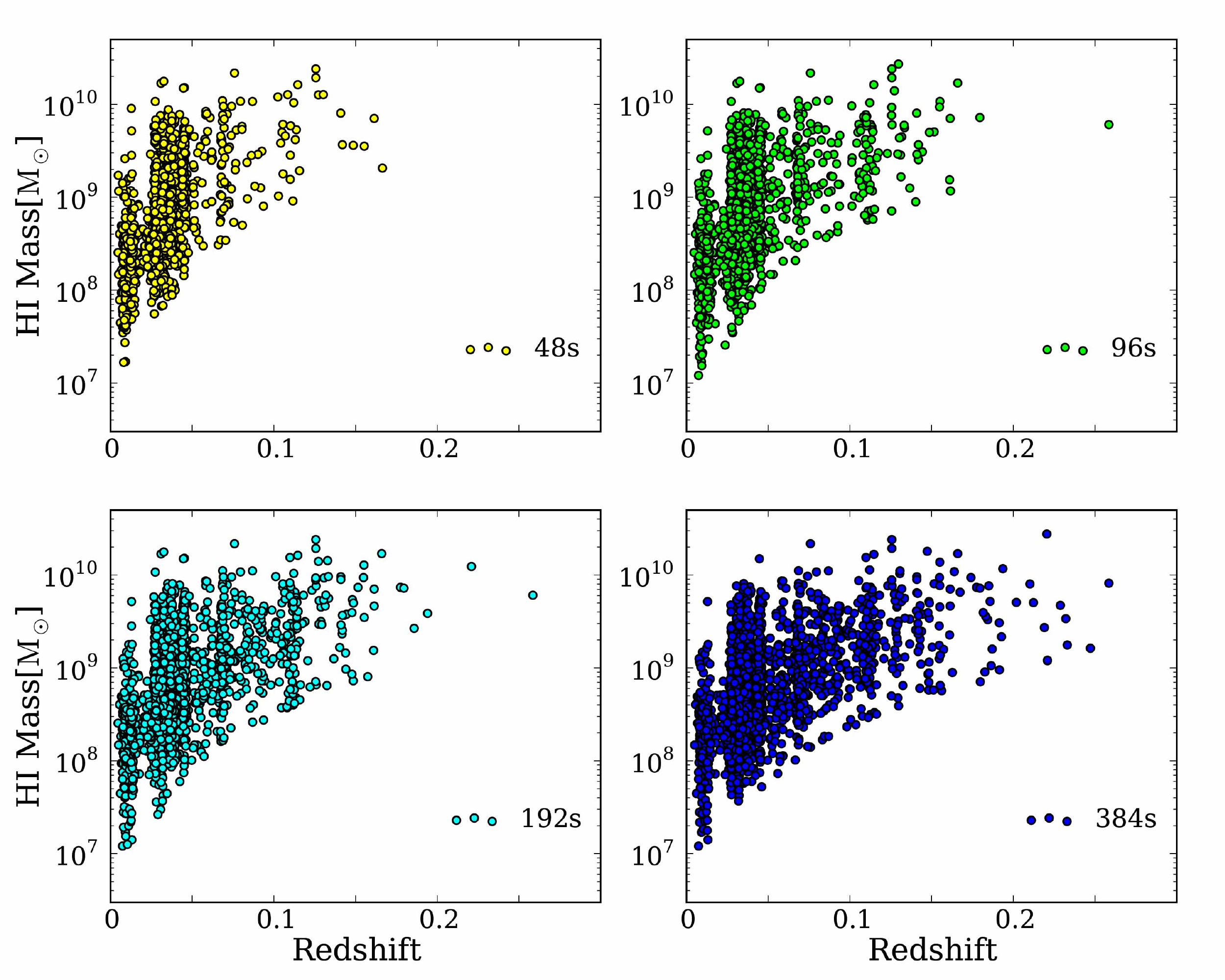}
  \caption{The distribution of the HI mass of the galaxy candidates from the 
  15$\times$15$\times$600 $h^{-3}\Mpc^3$ comoving volume.}
  \label{mass_dist_distribution_multitime}
\end{figure}

\begin{figure}
  \centering
  \includegraphics[width=0.4\textwidth]{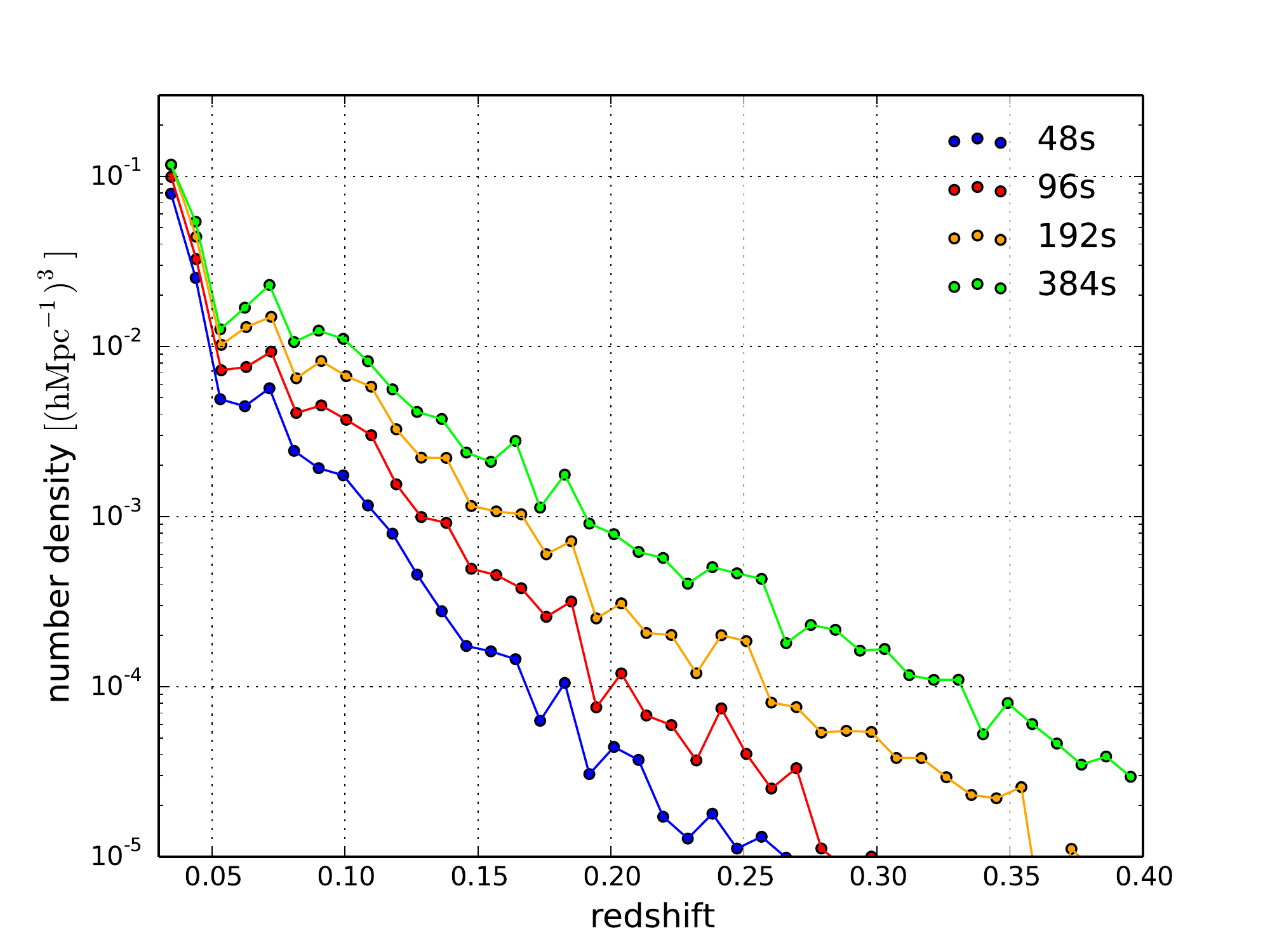}
  \caption{The comoving number density computed from the detected catalog.}
  \label{number_density_multitime}
\end{figure}

We plot the 10 $\Mpc/h$ slab of the mock galaxies detected in a two year (or 192s-integration-time) survey in 
Fig.~\ref{polar_distribution}, all galaxies in the simulated are also plotted for comparison. The masses and redshift of these detected galaxies are plotted in Fig.~\ref{mass_dist_distribution_multitime}, and in Fig.~\ref{number_density_multitime} we show 
the number density of the detected galaxies.
We can see from these figures that as redshift increases, the galaxies thinned out, and 
the number density decrease drastically as redshift increases. At $z=0.2$, the number density already falls off by two or three orders of magnitude compared with the nearby galaxies, and only the most 
massive galaxies can be detected at the higher redshifts.  Indeed, 
existing HI galaxy surveys are all limited to $z<0.2$, and even FAST, 
the largest single dish telescope in the world,  could not detect many galaxies in this mode. 
In the case of integration time of 48 s, the galaxy number density would drop to about $10^{-4} (\Mpc/h)^{-3}$ at $z=0.15$, 
corresponding to roughly the galaxy number density required for BAO measurement.
For the longer integration time, the distribution of the detectable galaxies extends to higher redshift. 
But even for the 384 s survey, the number density drops to $10^{-4} (\Mpc/h)^{-3}$ at $z=0.4$.

\begin{figure}
  \centering
  \includegraphics[width=0.4\textwidth]{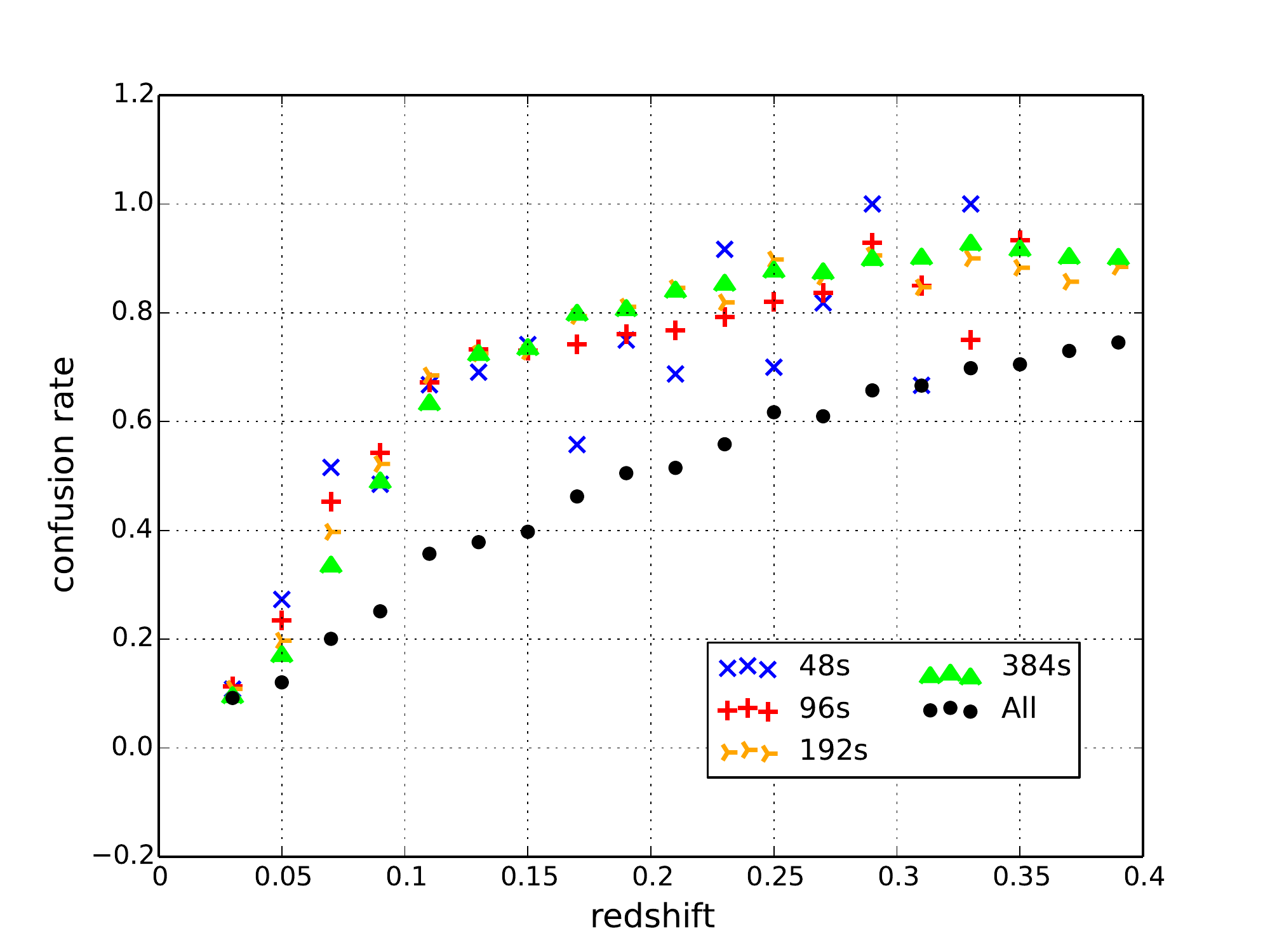}
  \caption{The confusion rate of the detected galaxies from different surveys. The minimum HI mass is set as $10^{8.5}M_{\odot}$.}
  \label{confusion_rate_ratio_multitime}
\end{figure}

For galaxy survey, galaxies may overlap with each other along the line of sight. The ability to uniquely identify the 
individual galaxy is an important evaluation of the performance of a telescope. To quantify this, we introduce the 
confusion rate, which is defined as the fraction of galaxies fall within the same voxel.
The voxel have a length of $0.08 \deg$ a side, which is 2 times of the FWHM of the FAST beam at $z=0$ , and a 
bandwidth of 1MHz in frequency axis. We have set a limit of HI mass of $10^{8.5}M_{\odot}$ in the 
mock catalog to remove the effect induced by low mass galaxies. 
For the galaxy detection algorithm  we used,  the confusion rate is shown 
in Fig. \ref{confusion_rate_ratio_multitime}. As the redshift increases, the confusion rate rises rapidly.
The integration time has a mild influence on the confusion rate, as the smaller galaxies are detected for longer
integration time, which leads to an increase in the number of galaxies per beam and hence a higher confusion rate.
This steady raise of the confusion rate shows how the galaxy count transits smoothly and naturally to 21cm intensity 
mapping with the increase of redshift in the FAST observation. 

For the detected galaxies, due to the finite beam size and the noise, there will also be errors in the measured positions.
To quantify this effect we compare the measured positions in the mock observation with the original position in the 
catalog. The measured central position is taken to be the flux weighted average, 
\begin{eqnarray}
    \nu_{c} = \frac{\int s(\nu)\nu d\nu}{\int s(\nu)d\nu}, \qquad \theta_c = \frac{\int s_{i}\theta_{i} d\theta}{\int s_{i}d\theta},
    \label{freq_c}
\end{eqnarray}
The frequency integration range covers about $\pm$200 km/s, with resolution of 0.0236MHz.  
We sum the flux from the sub-data cube along the frequency axis, the angular beam average is computed 
on a grid with spacing of 0.0133$\deg$, and beam width 0.08$\deg$. 

\begin{figure*}
      \centering
      \includegraphics[width=0.7\textwidth]{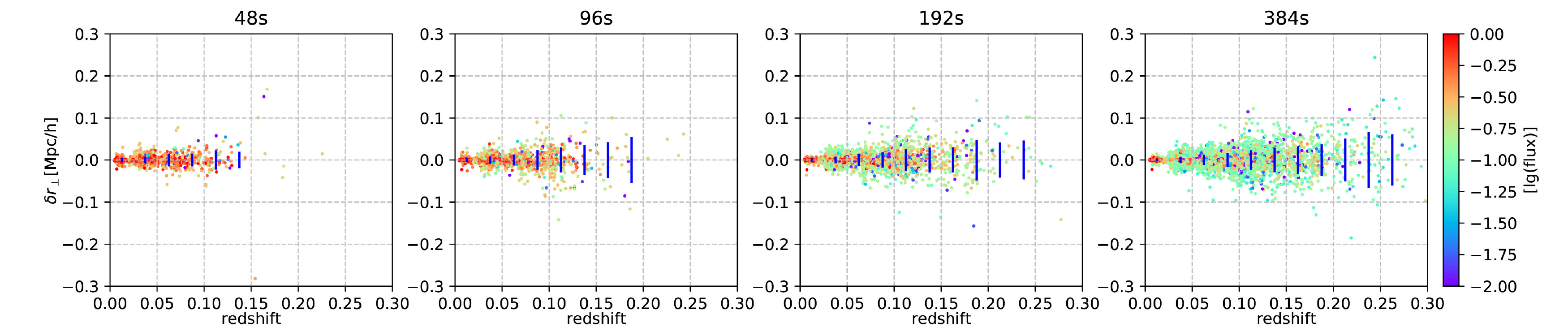}
      \includegraphics[width=0.7\textwidth]{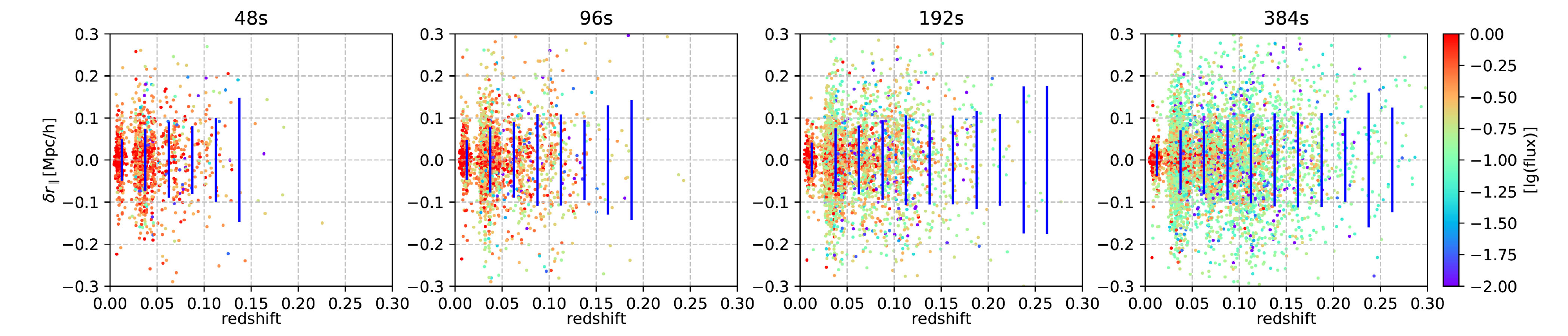}
      \caption{The position error induced by thermal noise in the observation for different integration times. Top panels: the 
      error perpendicular to the los. Bottom panels: the error along the los.  The galaxies are 
      colored according to their integrated flux.}
      \label{position_shift}
\end{figure*}

The error of the positions measurement in comoving coordinates is shown in Fig. \ref{position_shift}. We see that the noise 
in the galaxy HI profile can induce a shift in its position. For the plotted redshift range ($z<0.3$), the shift in the direction perpendicular to the line of sight (los) is generally smaller than that along the los direction. About 95$\%$ galaxies have position shifts in the perpendicular direction lower than 0.1$\Mpc/h$ and about 90$\%$ galaxies have shift in parallel direction lower than 0.5$\Mpc/h$. We colored the galaxies with their velocity-integrated flux. It shows that the galaxies with low velocity-integrated flux tend to have large shift. There are more galaxies which have large shift in survey with higher integration time, because more galaxies with low velocity-integrated flux can be detected with longer integration time. 
We also show the standard deviation of the position shift with blue vertical lines at different redshifts in 
Fig. \ref{position_shift}. Higher redshift has larger shift because there are more galaxies with lower velocity-integrated flux.

\begin{figure*}
  \centering
  \includegraphics[width=0.8\textwidth]{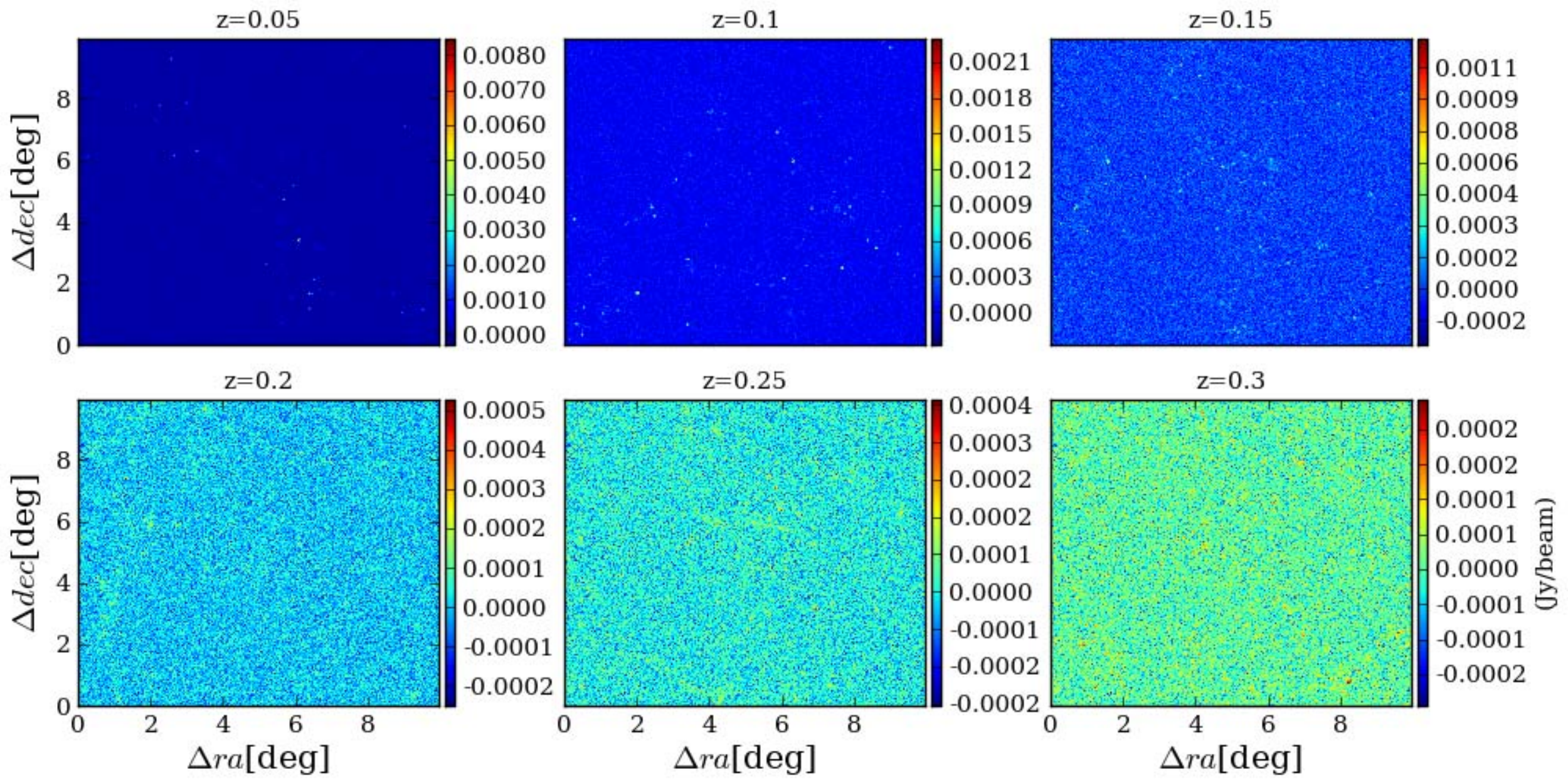}
  \caption{Noise-filled HI flux slice from different redshifts, all have the same frequency interval 1 MHz. The galaxy flux and noise both are computed with a bandwidth of 1 MHz and a integration time of 192s per beam.}
  \label{HI_intensity_flux_noise}
\end{figure*}
\begin{figure*}
  \centering
  \includegraphics[width=0.8\textwidth]{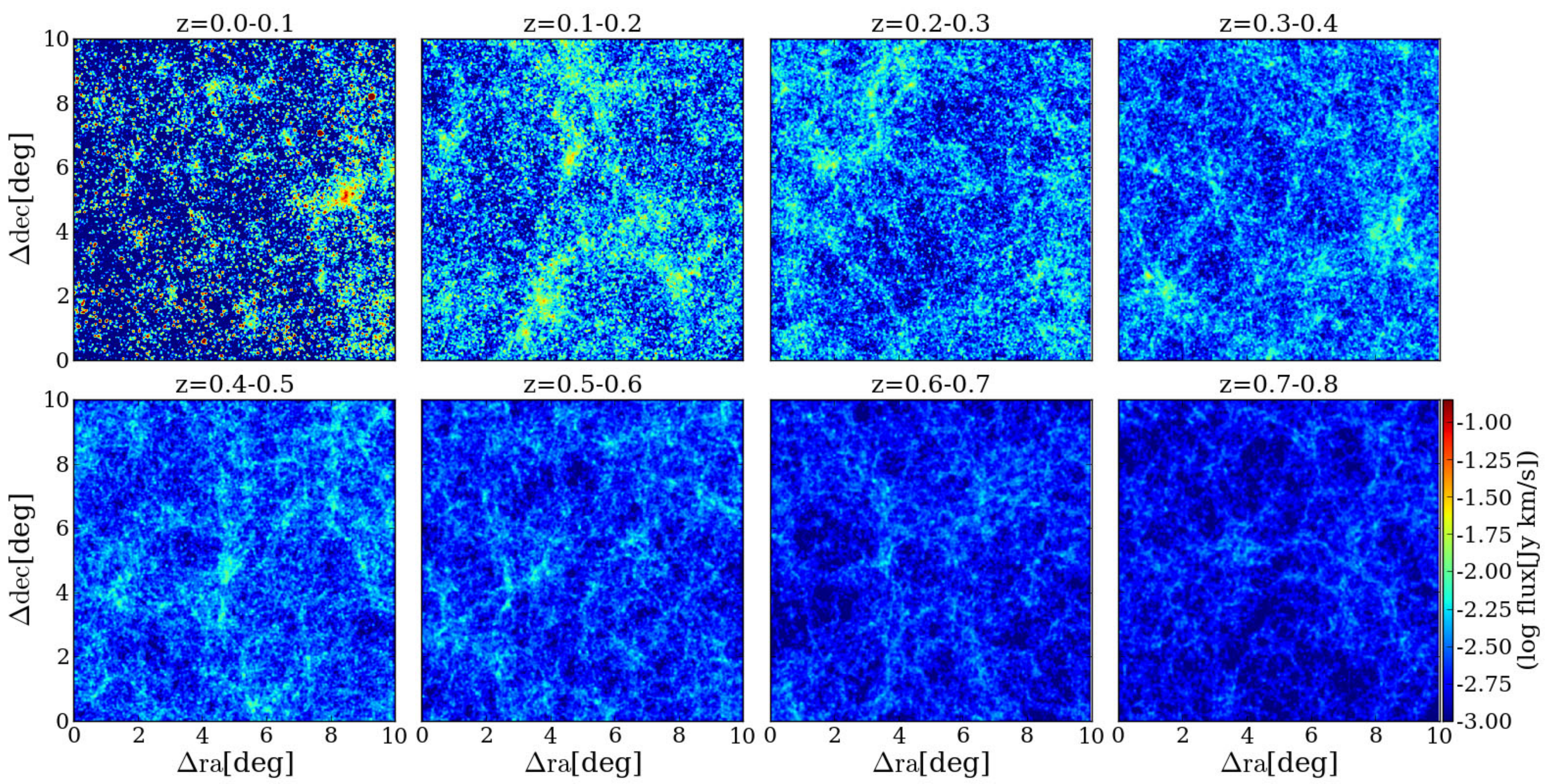}
  \caption{Velocity integrated HI flux map from different redshifts with $\Delta z = 0.1$.}
  \label{HI_intensity}
\end{figure*}

\subsection{HI Intensity Map}
In Fig.~\ref{HI_intensity_flux_noise} we show slices of HI intensity maps at six different redshifts in the range $0<z<0.3$, the depth is 1 MHz for each slice. We also added a thermal noise corresponding to 1 MHz of channel bandwidth 
and 192 seconds of integration time. To obtain the noise-filled maps, we construct the flux cube and noise cube with a bandwidth of 1 MHz, respectively, then combine them together. As we have mentioned above, missing the galaxies with HI mass below  $10^8 M_{\sun}$ will have a significant 
influence in intensity mapping experiment. We correct this by scaling all HI masses in the mock catalog with a z-dependent
 factor (> 1) to match $\Omega_{\rm HI}$ inferred from MUFASA cosmological hydrodynamical simulation. 
As can be seen from this figure, even for the relatively long integration time of 192 seconds, 
the map is still largely dominated by noise,  showing the challenges one will face in HI surveys.

In Fig.~\ref{HI_intensity}  we show the projected HI intensity maps at $0<z<0.8$, each with a redshift interval of 0.1.
The HI distribution is shown more clearly in these maps without the noise. For slices nearby, one can clearly see individual galaxies. As the distance increases, the structures become more 
blurred, and also the intensity drops. Individual galaxies become increasingly
difficult to see, but the overall structure remains, which illustrates how the intensity mapping could be used to probe the large scale structure. Note that the equal spacing in redshift means
slightly smaller comoving distance spacing at higher redshift, but we have checked and found that the equal spacing 
in comoving distance generate maps pretty similar to these.

\section{Power Spectrum Measurement}
The power spectrum is the most widely used statistics for large scale structure. In this section we describe its measurement and 
error forecast for both the HI galaxy survey and the HI intensity mapping survey with FAST.

\subsection{HI Galaxy Power Spectrum}

In a galaxy redshift survey with negligible error on the position of galaxies, 
the measurement error of the power spectrum comes from sample variance as well as shot noise.
Over a k bin of width of $\Delta k$ is \citep{1994ApJ..426...23F,2008MNRAS.383..150D}
\begin{eqnarray}
\frac{\sigma_{P}}{P} = \sqrt{2\frac{(2\pi)^3}{V_{\rm eff}}\frac{1}{4\pi k^2 \Delta k}}\frac{P(k)+1/n}{P(k)},
\label{projectederror_galaxy}
\end{eqnarray}
where
$V_{\rm eff}(k) = \int{\bigg[\frac{n(\vec{r})P(k)}{n(\vec{r})P(k) + 1}\bigg]^2}d^3\vec{r}$ for detected galaxies $n(r)$.
The error on galaxy position may also induce slight errors, but for the scale of interest they are negligible.

\begin{figure}
  \centering
  \includegraphics[width=0.4\textwidth]{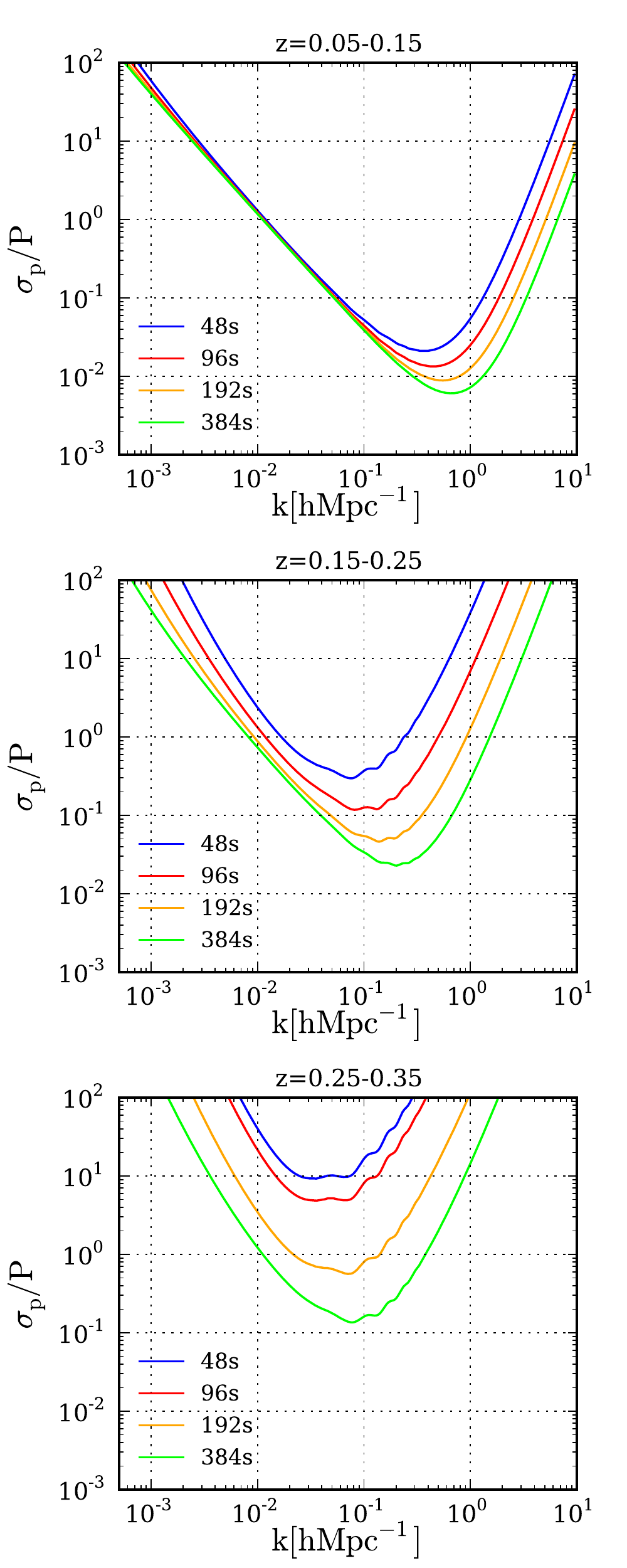}
  \caption{The projected error on power spectrum from a $20000\deg^2$ galaxy survey with integration time of 48 s, 96 s, 192 s and 384 s per beam. 
  At k$\approx$ 0.07 $h/\Mpc$ the S/N can reach 5.0 at $z\approx0.2, 0.25, 0.3$ and 0.35 respectively.} 
  \label{projected_error_galaxy_count}
\end{figure}

An optimal weighted estimator (known as the Feldman-Kaiser-Peacock or FKP estimator)
may be formed to minimize the measurement error of the power spectrum \citep{1994ApJ..426...23F}.
To make the measurement  in the irregular geometry of an actual survey, a mock sample of random points are generated. 
The detected galaxies are re-gridded into a rectangular box, 
the FKP estimate for the weighted density field is
\begin{eqnarray}
F(\bm{r}) = \frac{w(\bm{r})}{N}[n_{g}(\bm{r}) - \alpha n_{s}(\bm{r})],
\label{F}
\end{eqnarray}
where $w(\bm{r})=1/(1+\bar{n}(\bm{r})P)$ is the FKP weight, $\bar{n}(\bm{r})$ is the selection function (i.e. mean density) at the position $\bm{r}$, and P is an prior estimate of 
the power $P(k_1)$ at the scale of interest $k_1$, $n_{g}(\bm{r})$ and $n_{s}(\bm{r})$ refer to the number density 
of the observed galaxy catalog and the random mock catalog respectively, $\alpha$ is the real-to-mock ratio:
$\alpha = \sum w(\bm{r}_{\rm real})/\sum w(\bm{r}_{\rm random})$.
The  normalization factor $N$ in Eq. (\ref{F})  is given by
\begin{eqnarray}
N^2 = \int d^3r\bar{n}^2(\bm{r})w^2(\bm{r})
= \beta\sum_{\rm random}\bar{n}(\bm{r}_{i})w^2(\bm{r}_{i}),
\label{normal}
\end{eqnarray}
where  $\beta$ is the unweighted ratio of number of galaxies in the real (in our case simulation) catalog to that in the random mock catalog. 
In order to reduce the shot noise, the mock catalog is always set to contain much more galaxies than the real catalog. 
Here we choose $\beta$ to be 0.02. The power spectrum can then be estimated from the Fourier Transform of the weighted over-density field
$\langle |F(\bm{k})|^2\rangle$. The measurement error on power spectrum are estimated with $\Delta k/k = 0.125 $, for which the 
error of different $k$-bins can be regarded as uncorrelated \citep{1998ApJ...495...29G,1999MNRAS.304..851M}.

Our simulation data cube is in a pencil-beam shape, the amplitude and shape of the power spectrum estimated from the
cubic grid are biased, $\langle |F(\bm{k})|^2\rangle = W(k) \hat{P}(k)$. To correct for this effect, we compute the window function for this survey geometry by  
producing two  set of random catalogs, one is distributed only in the pencil-beam region, the other one 
in a cube region which encloses the pencil-beam region, the window function is then the ratio of the two power spectrum 
$P_{\rm pencil}(k)/P_{\rm cube}(k)$. We produce 10 pairs of samples and use the mean value to make the estimate. The true power spectrum is then obtained 
by dividing the window function. In summary, 
the power spectrum is obtained with the following steps:
\begin{enumerate}
\item Compute the selection function $\bar{n}(\bm{r})$ in each redshift bin. 
\item Produce the weighted over-density field $F(\bm{r})$) in the gridded box with grid spacing $1.0 \Mpc/h$ using the Nearest Grid Point assignment technique. 
\item Fourier transform the weighted over-density field, and compute the power spectrum $\langle |F(\bm{k})|^2\rangle$ with $\Delta k/k$ = 0.125. 
\item Correct the shape effect using the final window function,   $\hat{P}(k)= \langle |F(\bm{k})|^2\rangle /W(k)$. 
\end{enumerate}

The L-band receiver can cover a redshift range of up to $z=0.35$. In Fig. \ref{projected_error_galaxy_count} we show 
the projected error on power spectrum from a 20000-$\deg^2$ galaxy surveys for redshift $0.05<z<0.15, 0.15<z<0.25, 0.25<z<0.35$,
with integration time 48s, 96s, 192s and 384s per beam. On larger scales, the measurement precision is limited by 
the available number of modes (cosmic variance), while on the smaller scale it is limited by the available number of galaxies per 
cell (shot noise).  The best relative error in the power spectrum is achieved somewhere at $10^{-2} h/\Mpc <k < 10^{0} h/\Mpc $. 
At higher redshift the optimal point shift towards larger scales (smaller $k$), as the probed volume increases and 
the observed galaxy number density decreases.  At the BAO scale k $\approx$ 0.07 $h/\Mpc$, the signal to noise ratio can 
reach 5.0 at $z \approx 0.2, 0.25, 0.3$ and 0.35 respectively.   so in our Fisher estimation we will use these redshift regions 
as the survey volume of the FAST HI galaxy survey.  As can be seen from the figure, at the lower redshift, and especially on small $k$ values
the difference in the integration time per beam does not make much difference, however for the high redshift and large $k$ values the increase of 
integration time can significantly improve the measurement precision.  

\begin{figure*}
  \centering
  \includegraphics[width=0.75\textwidth]{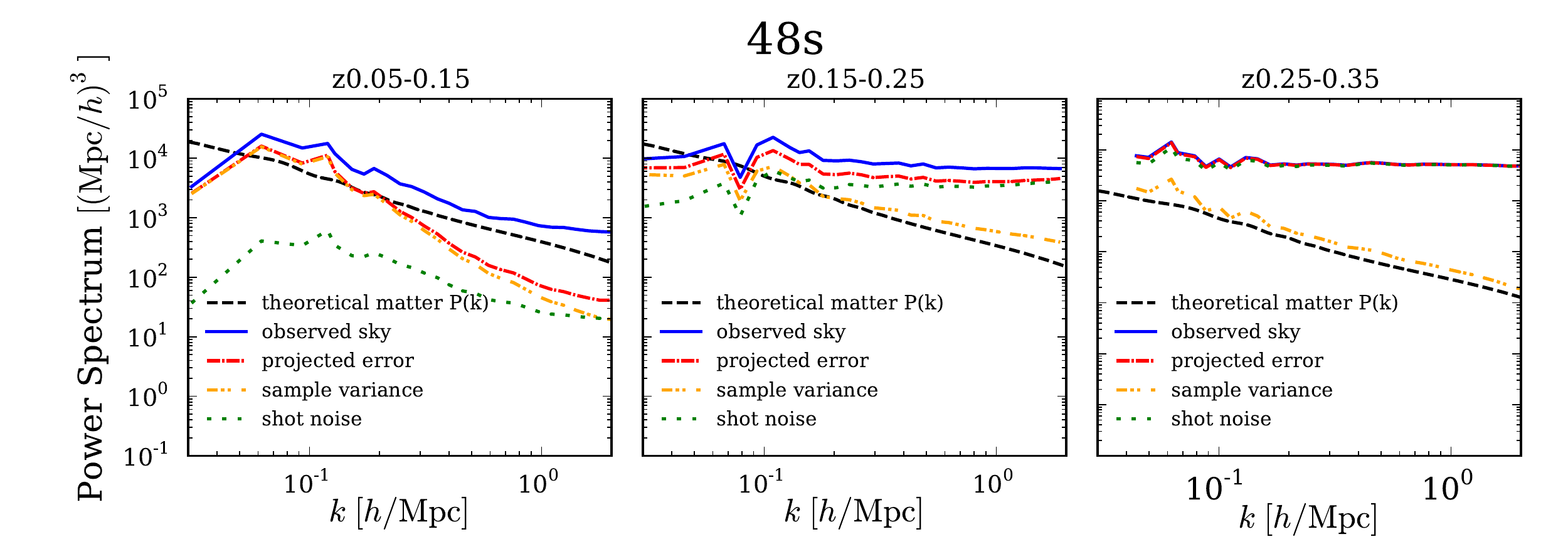}
  \includegraphics[width=0.75\textwidth]{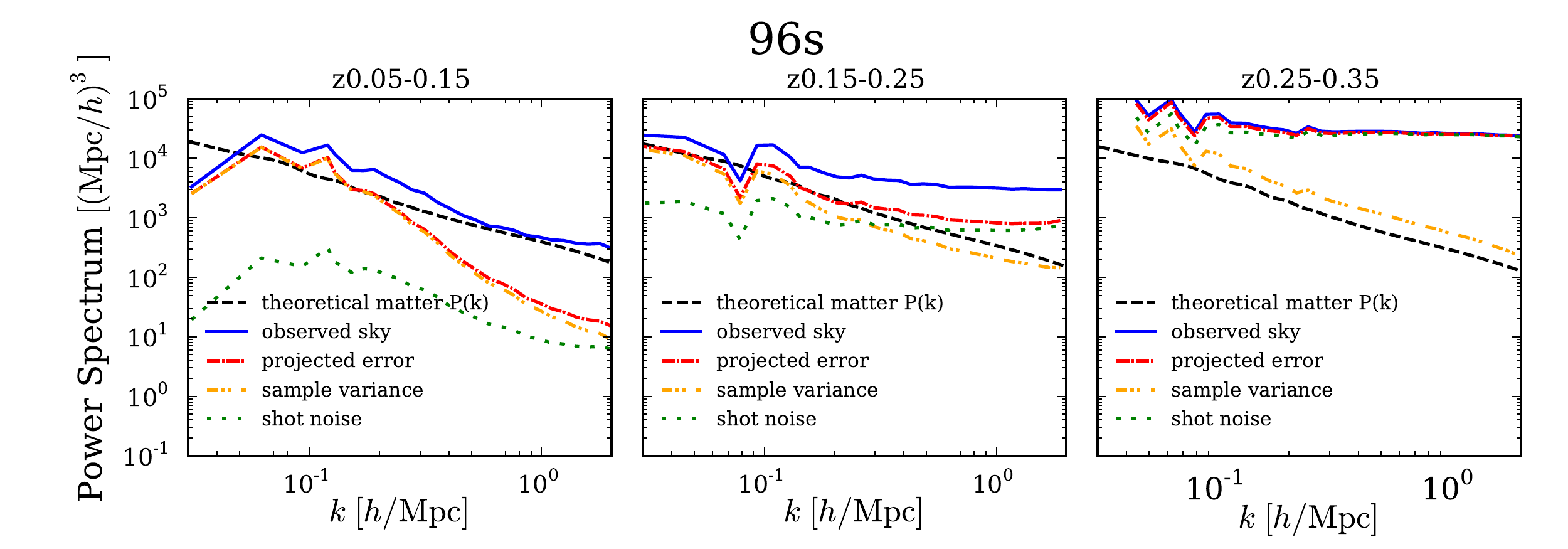}
  \includegraphics[width=0.75\textwidth]{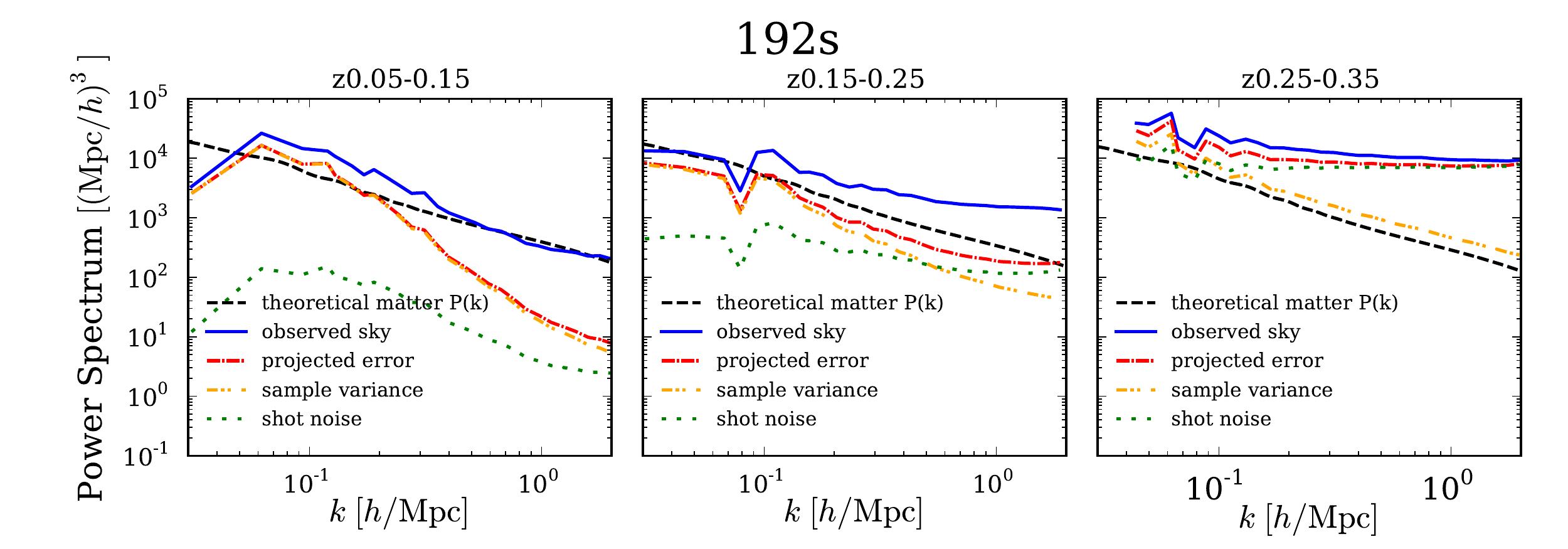}
  \includegraphics[width=0.75\textwidth]{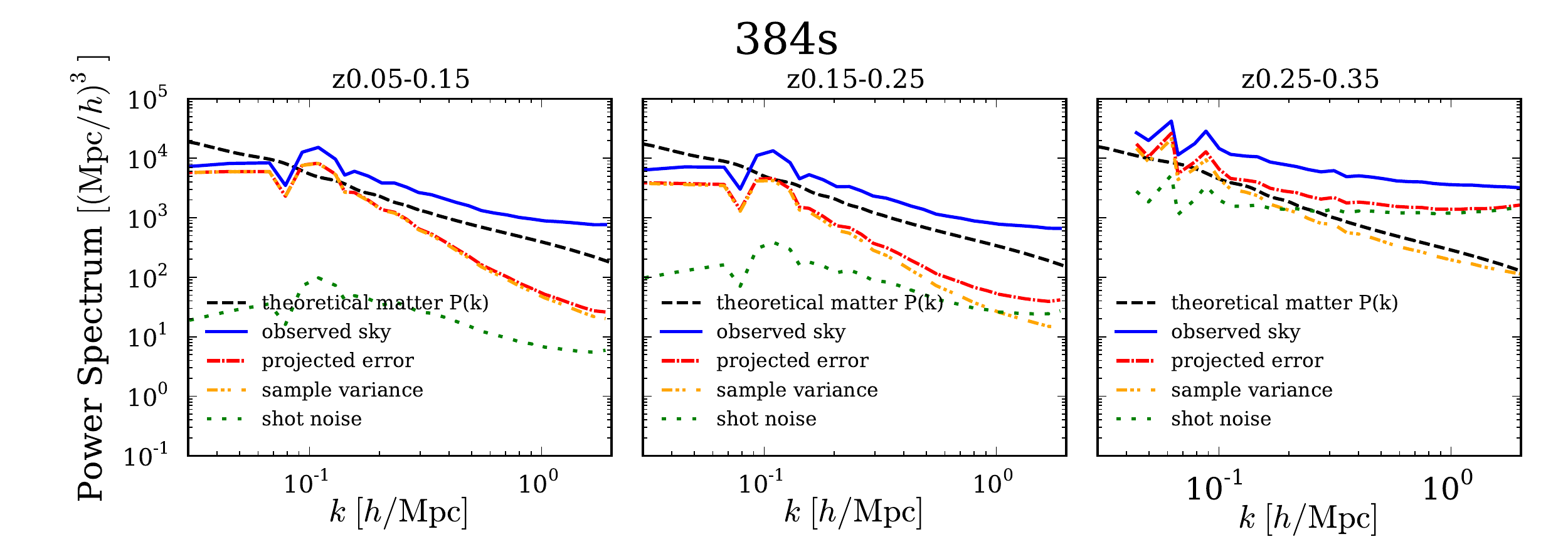}
  \caption{The HI galaxy power spectrum from simulation. We show the theoretical matter power spectrum 
  (black dashed line),  the simulated measurement (blue solid line), the total projected error estimated with Fisher matrix 
  (red dash-dotted line) and the error due to sample variance (orange dash-dot-dotted line) and shot noise (green dotted line). }
  \label{fkp_multitime}
\end{figure*}

The power spectra from the simulated observations are shown in Fig. \ref{fkp_multitime}. 
From top to bottom panel are the result with 48s, 96s, 192s and 384s respectively, each in the redshift range of 0.05-0.15 (left), 
0.15-0.25(middle), 0.25-0.35(right). In each subfigure, the theoretically expected power spectrum, the mock observation power spectrum, 
and the projected measurement error are plotted. For the noise, we also show separately the sample variance (corresponding to the size of the simulation
box) and the shot noise components. In this case, the cosmic variance is much larger as the size of simulation box 
spans  only a region of $10\times10\deg^2$.  

In this simulated galaxy survey, in almost all cases the measured power spectrum is larger than the theoretical power spectrum, 
as can be seen from Fig. \ref{fkp_multitime}. However, it does agree well with the projection. 
On small scales the shot noise dominated, which results in a nearly flat spectrum. This is particularly obvious at higher redshifts 
and for shorter integration times, where the number density of the detected galaxies is too small.  On larger scales, at lower redshifts 
and longer integration times, the shape of the power spectrum is more similar to the theoretical power spectrum, but there is still 
significant difference, and there are some large fluctuation at the large scales, thanks to the contribution of the cosmic variance. 
The overall normalization of the power is higher than the matter power spectrum (marked as theoretical), due to the fact that 
only the rare massive galaxies can be detected, and a clustering bias is introduced. The projected errors get larger in higher redshifts. 
However, we note that for a real survey with large sky area, the cosmic variance can be significantly reduced.

\subsection{Intensity Map Power Spectrum}

\begin{figure*}
  \centering
  \includegraphics[width=1.0\textwidth]{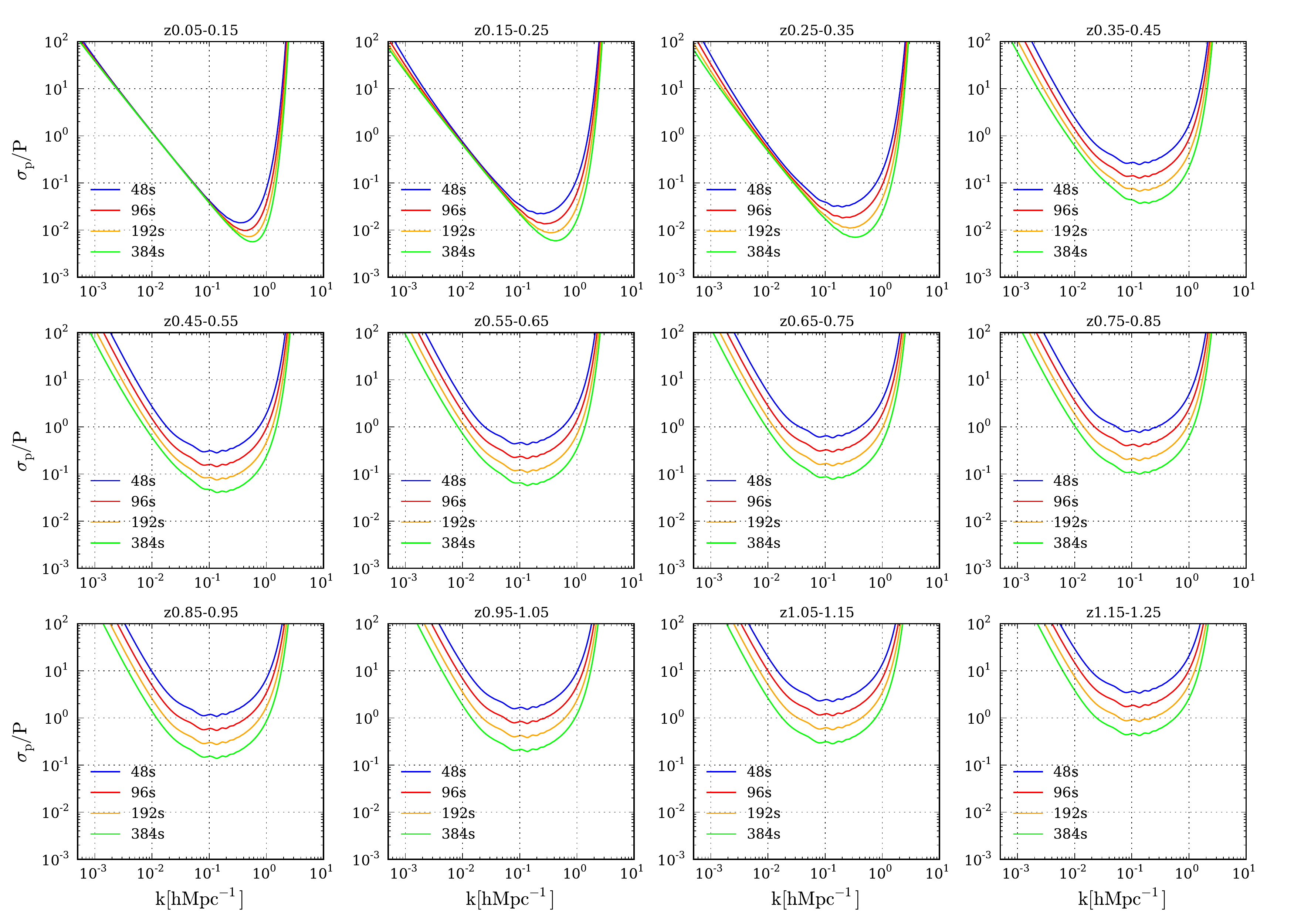}
  \caption{The projected error on power spectrum from FAST 20000 $\deg^2$ intensity mapping experiment with integration time of 48s, 96s, 192s and 384s per beam. At k $\approx$ 0.07 $h/\Mpc$ the S/N can reach 5.0 until redshift of 0.35, 0.55, 0.75 and 1.05 respectively.}
  \label{projected_error_im}
\end{figure*}

For intensity mapping survey, the measurement error of the power spectrum
can be written as \citep{2010ApJ...721..164S}

\begin{eqnarray}
    \frac{\sigma_{P}}{P}  & =  & 2\pi\sqrt{\frac{1}{V_{\rm eff}(k)k^2\Delta k}}, 
    \label{projectederror_im}
\end{eqnarray}
with the $V_{\rm eff}$ given in this case by
\begin{eqnarray}
V_{\rm eff}(\vec{k}) = V_{\rm sur}\left(1+\frac{\sigma_{\rm pix}^2V_{\rm pix}}{[\bar{S}(z)]W(\vec{k})^2P}+\frac{1/\bar{n}}{P}\right)^{-2}.
\label{Veff_im}
\end{eqnarray}
where $\bar{S}(z)$ is the average 21-cm emission flux density, $V_{\rm pix}$ is the pixel volume. 
The first term is due to sample variance, the second term is induced by the system thermal noise, 
and the last term is 
the shot noise due to the discreteness of the HI sources, with $1/\bar{n} \approx$ 100$h^{-3}\Mpc^3$ (\citet{2010ApJ...721..164S}). 
We model the angular resolution (the frequency resolution is much higher) as:
\begin{eqnarray}
W(k) = \exp\bigg[-\frac{1}{2}k^2r(z)^2\bigg(\frac{\theta_{\rm pix}(z)}{2\sqrt{2\ln{2}}}\bigg)^2\bigg]
\label{window}
\end{eqnarray}

\begin{figure*}
  \includegraphics[width=0.83\textwidth]{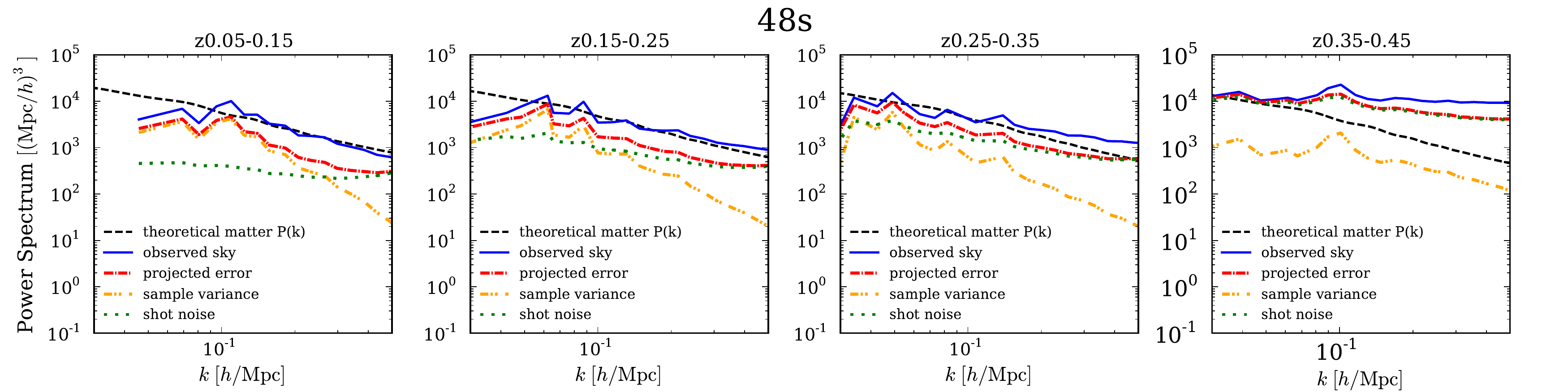}
  \vspace{-0.5ex}
  \includegraphics[width=0.83\textwidth]{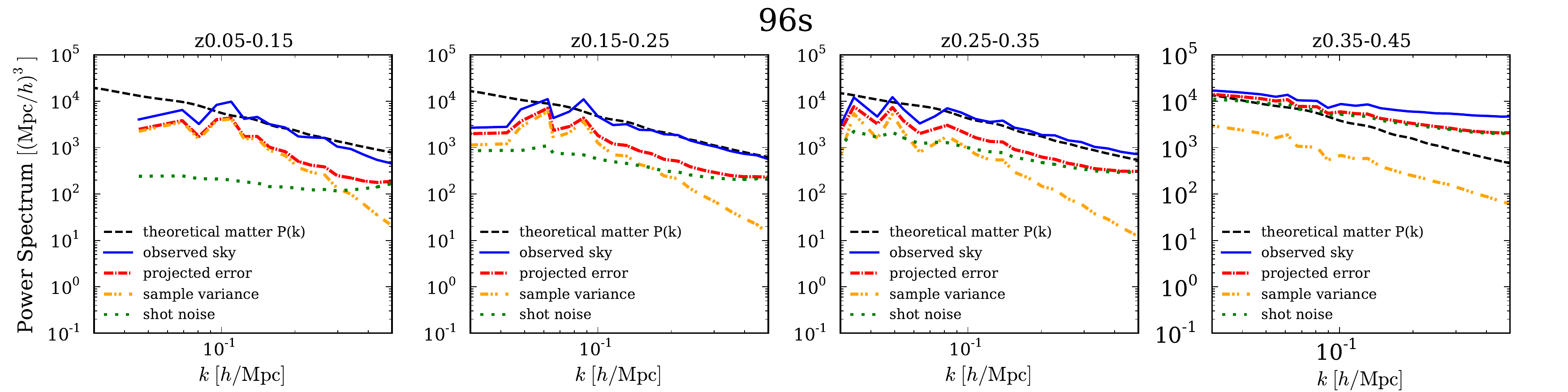}
  \vspace{-0.5ex}
  \includegraphics[width=0.83\textwidth]{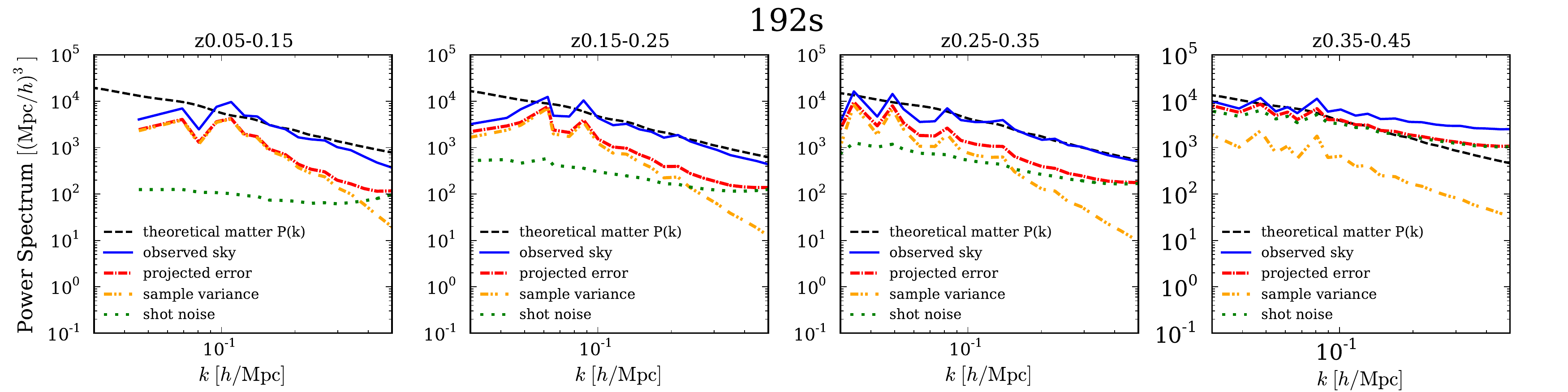}
  \vspace{-0.5ex}
  \includegraphics[width=0.83\textwidth]{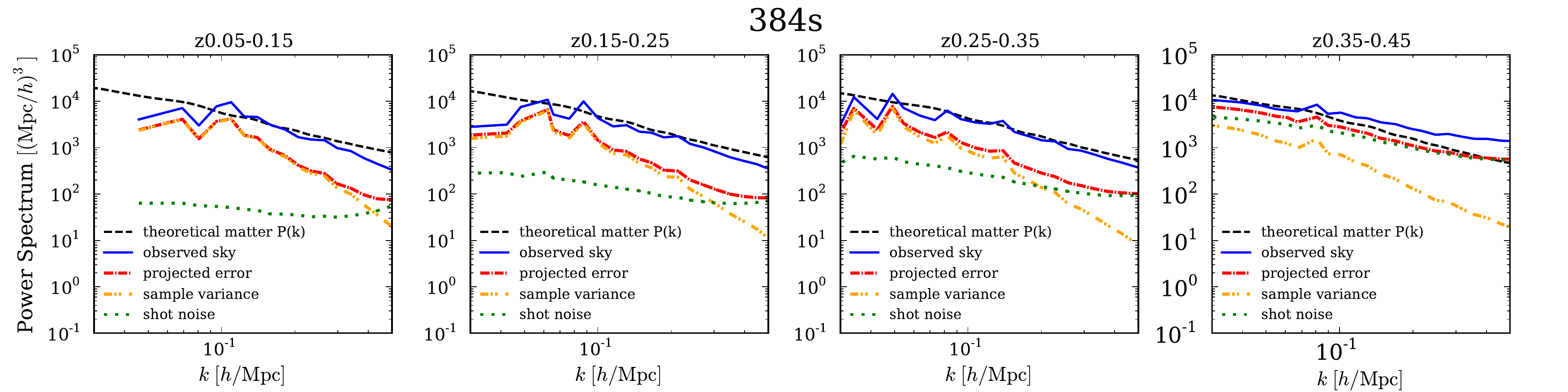}
  \caption{The intensity mapping power spectrum for different integration time per beam from simulation.}
  \label{intensity_mapping_multitime}
\end{figure*}

The image cube is gridded into a rectangular box with grid spacing of 2 $\Mpc/h$,  here we take $2 \Mpc/h$ sized pixels as the 
standard sampling size, it is well below the BAO scale which is about $150 \Mpc/h$ and would not affect the cosmological result.  
The HI emission flux density field is related to the over density field by:
\begin{eqnarray}
\delta(\bm{r}) = S_{\rm HI}(\bm{r})/\bar{S}_{\rm HI}- 1,
\label{flux_to_density}
\end{eqnarray}
where $S_{\rm HI}(\bm{r})$ is the HI emission flux density at position $\bm{r}$ and $\bar{S}_{\rm HI}$ is the mean flux density of HI emission. 
We also add a Gaussian thermal noise of  0.74mJy, 0.52mJy, 0.37mJy and 0.26mJy per beam, corresponding to 48s, 96s, 192s and 384s 
integration time per beam. The simulated power spectrum is measured as follows:
\begin{enumerate}
\item Generating the mock data filled with Gaussian thermal noise,
\item Re-gridding the image data cube to rectangular box with grid spacing of 2 $\Mpc/h$,
\item Converting the flux density field to HI mass overdensity field, 
and Fourier transform the field to get the power spectrum $\langle |F(\bm{k})|^2\rangle$ with $\Delta k/k$ = 0.125.
\item Using the pencil beam survey window function to correct the shape effect.
\end{enumerate}

The statistical error is estimated as follows:
\begin{enumerate}
    \item Generating a synthetic product with a spatial resolution same as FAST beam size and a frequency resolution of 1MHz, S($r_{\rm x}$,$r_{\rm y}$,z),
    \item Generating a mock data filled with Gaussian thermal noise, N($r_{\rm x}$,$r_{\rm y}$,z), the noise data cube have the same size and 
    resolution as S($r_{\rm x}$,$r_{\rm y}$,z),
    \item Re-gridding the synthetic data S($r_{\rm x}$,$r_{\rm y}$,z) and the noise data N($r_{\rm x}$,$r_{\rm y}$,z) to a rectangular box with grid spacing of 2 $\Mpc/h$, obtaining S($r_{\rm x}^{\prime}$,$r_{\rm y}^{\prime}$,z) and N($r_{\rm x}^{\prime}$,$r_{\rm y}^{\prime}$,z), 
    \item Computing the mean signal of S($r_{\rm x}^{\prime}$,$r_{\rm y}^{\prime}$,z) and the variance of N($r_{\rm x}^{\prime}$,$r_{\rm y}^{\prime}$,z) at different redshift, obtain $\bar{S}(z)$ and $\sigma_{\rm pix}$, 
    \item Obtaining the projected error by use of Eq. (\ref{projectederror_im}).
\end{enumerate}

Figure. \ref{projected_error_im} shows the statistical error of the power spectrum that can be achieved by $20000 \deg^2$ FAST intensity 
mapping survey with integration time 48s, 96s, 192s and 384s per beam respectively. Here, we ignore for the moment the frequency range
of the L-band receiver system, but simply plot up to much higher redshifts, which may be accomplished with the wide-band single feed receiver, or a 
future UHF PAF receiver introduced in Section 2.1. For simplicity and easier comparison, here we also plot the projected error for the same integration 
time, though to the size of the beam and the number of available beams the required total survey time would be very different at the higher redshifts.

A comparison with Fig.~\ref{projected_error_galaxy_count} shows that for the lowest redshift bin ($0.05<z<0.15$), the projected errors 
are almost the same as the HI galaxy survey of the same integration time. We saw from Fig.~\ref{number_density_multitime} that even in this 
low redshift range, the comoving number density of the detected HI galaxies are decreasing with increasing redshift, showing that the FAST does not
detect all HI galaxies. Nevertheless, the detected HI galaxy has sufficiently high number density that it gives a good representation of the underlying 
HI galaxy sample and total mass density. However, for all the other redshift bins the projected error of the 
intensity mapping is much smaller than the HI galaxy survey. In fact, as redshift increases, the signal-to-noise ratio 
actually improves for a while as the survey volume increases and sample variance decreases, though 
eventually it begins to drop as the thermal noise becomes higher and HI signal becomes weaker.  
The shift in the optimal comoving scale (the minimum of the relative error in the power spectrum) at different redshifts are also 
much less than the HI galaxy surveys. 
At the BAO scale $k \approx$ 0.07 $h/\Mpc$, the S/N can reach 5.0 until
redshift of 0.35, 0.55, 0.75 and 1.05 for survey with 48s, 96s, 192s and 384s respectively. 

The simulated power spectrum measurement is shown in Fig. \ref{intensity_mapping_multitime}. The left three columns show the simulation with 
the same L-band receiver, while the column on the right shows a higher redshift ($0.35<z<0.45$) with the wide-band receiver system of a single feed. 
Compared with Fig.~\ref{fkp_multitime}, the thermal noise is replaced by the shot noise at the smaller scales.  Within the L-band (up to $z=0.35$), 
it is sub-dominant even for the shortest integration time considered here (48 s). So the intensity mapping can yield much nicer results than galaxy surveys on 
the small scales, and good precision can be achieved at  higher redshifts. On the larger scales where the cosmic variance dominates, 
the precision of this simulated measurement is limited by the size of the simulation box so we see some large fluctuations, 
but this can be significantly reduced with larger volumes.  The intensity mapping survey can efficiently map the  large-scale structure of 
BAO scale until $z \approx 0.35$ even with moderate integration time. Note that the error from intensity mapping is 
smaller at most of scales, but at small scales ($k \approx 1.0 h/\Mpc$) the noise is larger than HI signal at all redshifts due to the limited 
resolution.

For redshift $0.35<z<0.45$ case, i.e. the right column in Fig.~\ref{intensity_mapping_multitime},  the wide-band receiver is assumed to be 
used as the corresponding frequency range is beyond that of the L-band receiver.  
This receiver has a much higher noise level, the system temperature is 60 K, so the survey precision for the 
same integration time degraded a lot. Also, as it is a single 
feed receiver, the required survey time for accomplishing the same integration time would be much longer. In order to make a good intensity mapping survey 
at the higher redshifts, one must be able to have multi-beam receivers, preferably with low system noise. In the next subsection where the cosmological 
measurement is discussed, we shall assume that such a receiver, namely the UHF PAF receiver introduced in Sec.2.1 will be available in the future for 
such surveys.

\subsection{Cosmological Constraints}

Because the limit of our survey volume, we estimate the induced measurement error on cosmological parameters by using the
Fisher matrix formalism \citep{1997PhRvL..79.3806T,2003ApJ...598..720S,2013MNRAS.430..747M}. The Fisher matrix for 
 parameter set $\{p_i\}$ is given by
\begin{equation}
F_{ij} = \int_{\vec{k}_{\rm min}}^{\vec{k}_{\rm max}}\frac{\partial \ln P(\vec{k})}{\partial p_{i}}\frac{\partial \ln P(\vec{k})}{\partial p_{j}}V_{\rm eff}(\vec{k})\frac{d\vec{k}}{2(2\pi)^3} 
\label{fisher}
\end{equation}
The usable range of $k_{\rm min}$ and $k_{\rm max}$ are assumed to be 
$k_{\rm min} = 10^{-3} h/\Mpc$ (from survey volume) 
and $k_{\rm max} = 0.1 h/\Mpc$. The observed power spectrum $P(\vec{k})$ is given by 
\begin{align}
    \begin{split}
    P_{\rm obs}(k_{\rm ref\perp},k_{\rm ref\parallel}) &= \frac{D_{A}(z)_{\rm ref}^2H(z)}{D_{A}(z)^2H(z)_{\rm ref}}b_{\rm HI}^2\left(1+\beta\frac{k_{\parallel}^2}{k_{\perp}^2+k_{\parallel}^2}\right)^2\\
                                                   &\hspace{1em}\times \left(\frac{G(z)}{G(z=0)}\right)^2P_{\rm m,z=0}(k) + P_{\rm shot},
    \end{split}
    \label{power_obs}
\end{align}
where $b_{\rm HI}$ is the linear bias factor of HI gas and the redshift space distortion factor $\beta = \Omega_{m}(z)^{0.6}/b_{\rm HI}(z)$. The Hubble parameter and the angular diameter distance can be computed for a model 
with dark energy equation of state parameterized in the form $ w(z) = w_{0}+w_{a}\frac{z}{1+z} $ \citep{2001IJMPD..10..213C}. 
To obtain useful constraints on cosmological parameters, it is necessary to break the degeneracy by combining the BAO data with data obtained from some other cosmological observations, e.g., CMB. The total Fisher matrix on distance parameters is given by 
\begin{equation}
F^{\rm tot} = F^{\rm CMB} + \sum_{i}F^{\rm LSS}(z_{i}), 
\label{fisher_tot}
\end{equation}
and $F^{LSS}(z_{i})$ is the Fisher matrix derived from the i-th redshift bin of the large scale structure survey. For FAST HI galaxy survey and intensity mapping observations, we divide the redshift region into several bins with equal redshift interval, which we set as 0.05. 

The HI gas mostly distributed in galaxies hosted by halos after the reionization, thus the HI bias can be modeled as the halo bias weighted by HI mass hosted by these halos:
\begin{align}
    b_{\rm HI} = \frac{1}{\rho_{\rm HI}(z)}\int_{M_{\rm min}}^{M_{\rm max}}dM\frac{dn}{dM}(M,z)M_{\rm HI}(M)b(M,z),
    \label{bias_HI}
\end{align}
where $dn/dM$ is the halo mass function, for which we use the fittings of (\citealt{2008ApJ...688..709T,2010ApJ...724..878T})
of the Sheth-Tormen function \citep{2002MNRAS.329...61S}.
The HI mass in a halo of mass $M$ is given by $ M_{\rm HI}(M) = A M^{\alpha}$ 
where the prefactor  $A$ will be canceled in the normalization of $\rho_{\rm HI}(z)$, 
and  $\alpha \simeq 0.6$ \citep{2015aska.confE..19S}. 
The halo bias in this ellipsoidal collapse model is modeled as,
$$
    b(M,z) = \frac{1}{\delta_{c}(z)}\bigg[\nu^{\prime2}+b\nu^{\prime2(1-c)}-\frac{\nu^{\prime2c}\sqrt{a}}{\nu^{\prime2c}+b(1-c)(1-c/2)}\bigg],
$$
where $a = 0.707$, $b = 0.5$, $c=0.6$, $\nu^{\prime} = \sqrt{a}\nu$, and  $\nu = \frac{1.686}{D(z)\sigma_{R}}$, 
where $D(z)$ is the linear growth factor \citep{2002MNRAS.336..112M}.

The bias for the HI galaxy survey and the HI intensity mapping survey is different. For the HI galaxy survey, the $M_{\rm min}$ is 
given by the minimum mass of detected galaxies in the survey, while for HI intensity mapping survey it is the minimum mass, about $\sim 10^6 \Msun$.

\subsection{Results}

We consider the constraint on dark energy EOS parameters ($w_{a}$,$w_{0}$) from the HI galaxy surveys and intensity mapping surveys. 
The cosmological constant point $w_0, w_a=(-1,0) $ is taken as the fiducial model. 
The L-band 19 beam receiver is assumed to be used for the redshift $0.05<z<0.35$. We then consider the cosmological constraint derived from (i) the 
L-band survey only; (ii) the L-band survey plus the existing wide-band single feed receiver; or (iii) the L-band plus the future
PAF receiver with 81-beams in the UHF band for higher redshift intensity mapping (see Sec.\ref{sec:instr} for descriptions). In each case, we consider 
the integration time per beam on the L-band is 48s, 96s, 192s, and 384s, respectively. As discussed in Sec.\ref{sec:instr}, the 48s integration can be 
completed in one scan of the 19-beam receiver.
When the wide band single feed receiver is used, for simplicity, we assume it acquires the same amount of integration time per beam. 
For the PAF in UHF band, we have considered instead two integration time per beam, 216 s corresponding to (one scan) or 432 s (two scans) of the 
PAF receiver. These are added to the L-band 192 s and 384 s respectively for illustration. 

These survey configurations are listed in the first column of Table \ref{covariance}. The precision of the dark energy EOS parameters $(\sigma_{w_0}, \sigma_{w_a})$ are given in the second column for galaxy survey and third column for 
intensity mapping survey. The required total observation time corresponding to each survey are 
given in the fourth column of the table, 
where the time required for each band are given separately.  
Note that the time listed is for observations, not counting offline time required for 
calibration, maintenance, etc., so the real time required to complete the survey would be even longer.  

\begin{figure}
  \centering
  \includegraphics[width=0.45\textwidth]{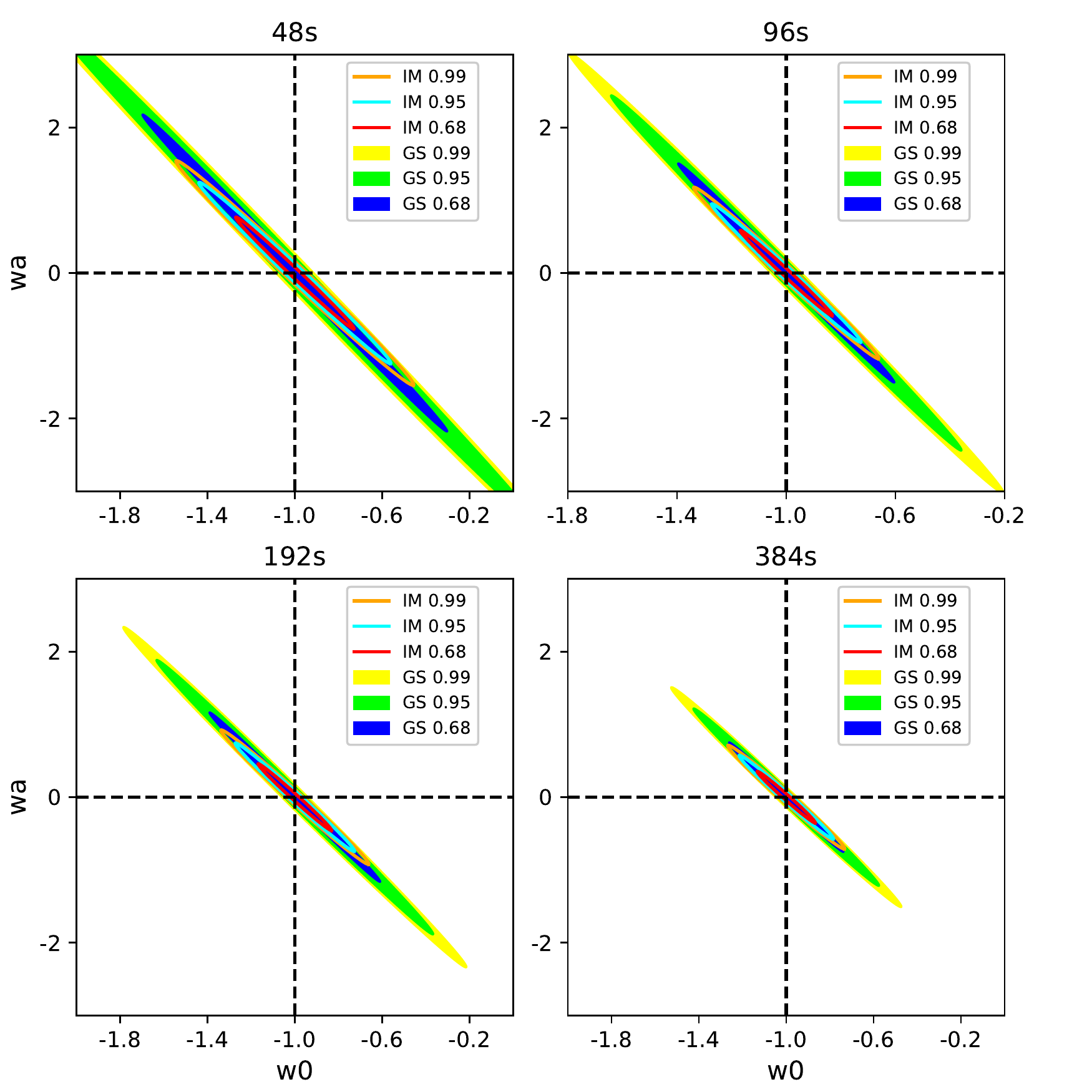}
  \caption{The constrains on dark energy EOS parameters from FAST galaxies survey and intensity mapping, both combined with Planck CMB observation. The three colors are iso-probability contours for 0.68, 0.95 and 0.99, respectively. Galaxies survey (GS) is labeled as filled ellipse and Intensity mapping (IM) is labeled with solid lines.}
  \label{w0_fld_wa_fld}
\end{figure}

\begin{figure}
  \centering
  \includegraphics[width=0.4\textwidth]{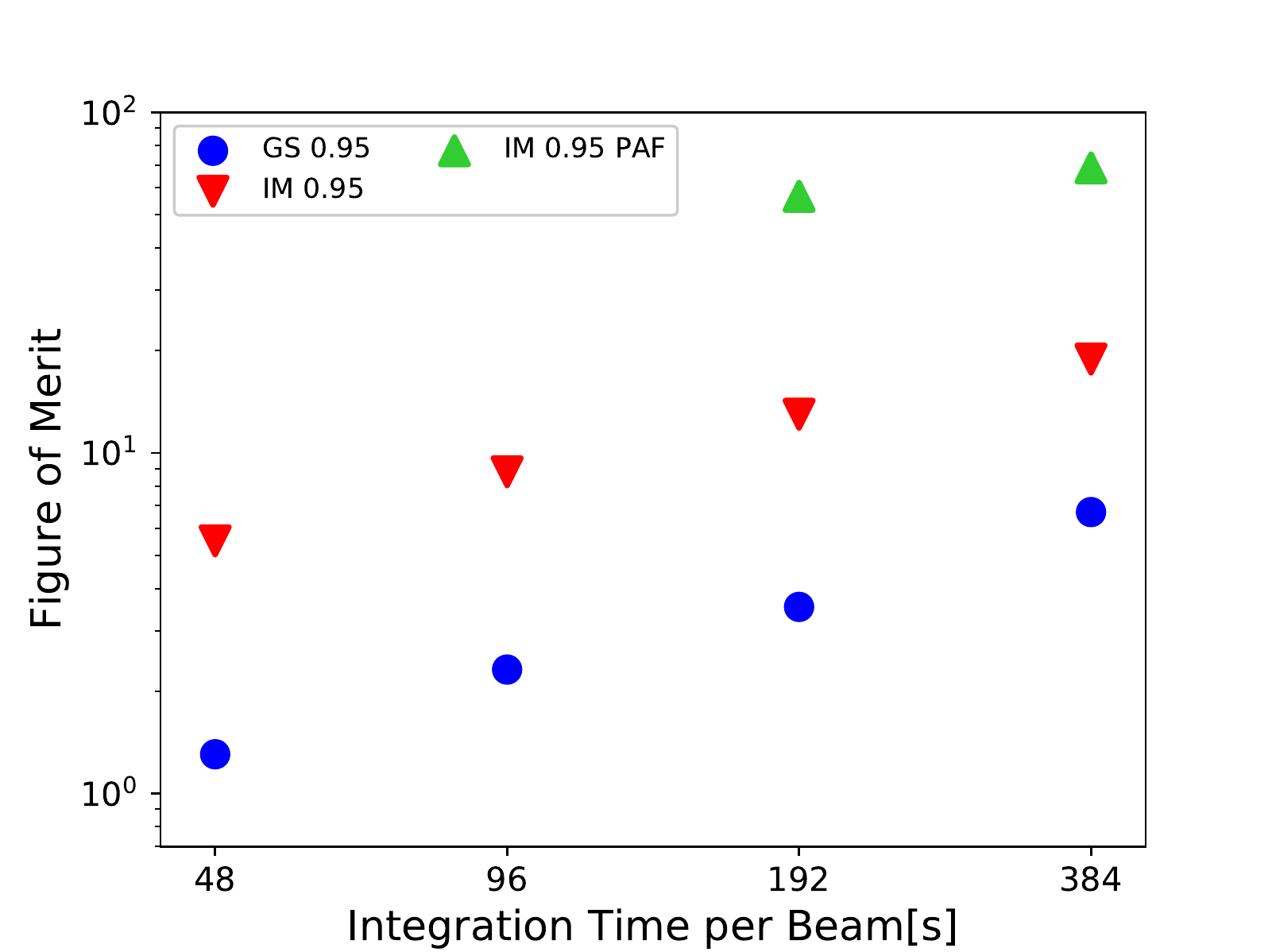}
  \caption{The figure of merit for dark energy EOS parameters from FAST galaxies survey and intensity mapping combined with Planck CMB observation. Galaxy surveys (GS) are labeled as filled blue circles and intensity mapping surveys (IM) are labeled with filled red down-pointing triangle, for confidence limit of 0.95. The IM surveys with low frequency PAF are shown in green up-pointing triangle symbols.}
  \label{w0_fld_wa_fld_FOM}
\end{figure}

We also show in Figure. \ref{w0_fld_wa_fld}  the error ellipses of the dark energy EOS parameters with the L-band 19 beam receiver and the wide band single 
feed receiver. The results for the L-band + PAF surveys are not shown as their error ellipses are much smaller.  
The DETF figure of merit, which is defined as the inverse of the area of the $2\sigma$ error ellipse \citep{2006astro.ph..9591A}, 
is shown in Figure. $\ref{w0_fld_wa_fld_FOM}$ for the L-band + wide-band receiver as well as the L-band + UHF-band PAF. 

\begin{table}
 \caption{Constraint on dark energy EOS parameters $(w_0, w_a)$ for FAST galaxy surveys (GS) and intensity mapping (IM) surveys  
 with a total of  $20000 \deg^2$ area and Planck prior. In first column the L denotes L-band 19 feed receiver, 
 w denotes wide band single feed receiver, and P denotes the PAF receiver in UHF band, and the number denotes 
 integration time in seconds per beam.  The last column shows the total time for the survey to be completed. }
 \label{covariance}
 \begin{tabular}{L{2cm}|C{1.4cm}|C{1.4cm}|L{2.3cm}}
 \hline
 Survey &GS $(\sigma_{w_0}, \sigma_{w_a})$ & IM $(\sigma_{w_0}, \sigma_{w_a)}$ & Observation Time (day)\\
 \hline
  L 48s & (0.46, 1.44)& (0.19, 0.53) &220\\
  L 96s & (0.33, 1.00)& (0.15, 0.43) &440\\
  L 192s & (0.25, 0.77)&(0.13, 0.36) &880\\
  L 384s & (0.17, 0.49)& (0.12, 0.33) &1760\\
 (L + w) 48s & (0.46, 1.44)& (0.18, 0.50) &220(L) + 2422(w)\\
 (L + w) 96s & (0.33, 1.00)& (0.14, 0.39) &440(L) + 4844(w)\\
 (L + w) 192s & (0.25, 0.77)& (0.11, 0.30) &880(L) + 9688(w)\\
 (L + w) 384s & (0.17, 0.49)& (0.09, 0.23) &1760(L) + 19376(w)\\
 L(192s) +P(216s) & - & (0.05, 0.12) &880(L) + 135(P)\\
 L(384s) +P(432s) &- & (0.04, 0.10) &1760(L) + 270(P)\\
 \hline
 \end{tabular}
\end{table}

From these we see the intensity mapping can achieve much higher precision in the measurement of the dark energy EOS parameters than the 
HI galaxy surveys. The intensity mapping survey with the shortest integration per beam (48s) has a figure of merit comparable with the 
HI galaxy survey of the longest integration time per beam (384s), but requires only 1/8 of the total observation time in the L-band. However, even with 
the intensity mapping survey, the figure of merit is only of order $10^1$, much less than the current optical surveys. This is not surprising as the L-band is
limited to relatively low redshifts ($z<0.35$). However,  we note that so far there has not yet been an HI survey providing constraints on the dark energy 
parameters, and the HI survey complements the optical survey as it uses a difference tracer, which would be valuable to reduce any possible 
systematic errors in the BAO measurement.

From Table \ref{covariance}, we also see that the constraints on the EOS parameters are only slightly improved by adding the surveys of equal integration
time per beam with the wide-band receiver. This is because this receiver has a higher system temperature (60 K) 
over its very wide frequency coverage. Furthermore, because it has only a single feed, to achieve such integration time per beam would require very long
observation time which are quite impractical. However, if equipped with a powerful PAF receiver, the measurement can be taken to much higher redshift 
in reasonable time--indeed, for the PAF parameter we assumed, it would take even less time to complete than the L-band. The figure of merit could then be 
lifted substantially, up to a level comparable with Dark Energy Task Force (DETF) stage IV experiments \citep{2006astro.ph..9591A}.  This shows that a
PAF receiver at the UHF band would be a very valuable addition to the FAST telescope.

\section{Discussions}
\subsection{The choice of galaxy survey vs. intensity mapping}
Using the Fisher information, \citet{2018arXiv180906384C} develop a formalism to quantify the performance of galaxy redshift survey 
and intensity mapping when measuring large-scale structures. Under the assumption that the galaxy population follows 
a Schechter function form, 
\begin{eqnarray}
n(L) &=& \phi_* \left(\frac{l}{l_*}\right)^{\alpha} e^{-l/l_*},
\end{eqnarray}
the optimal strategy for survey can be found using the relative value of three parameters: 
$\{L_{\rm SN}$, $\sigma_{\rm L}$, $l_{\ast}\}$, where $L_{\rm SN}$ is the luminosity scale on which the voxels 
are susceptible to shot noise, $\sigma_{\rm L}$ refers to the rms noise per voxel, and $l_{\ast}$ is the characteristic luminosity. 
The $L_{\rm SN}$ is derived with $\sigma_{\rm SN}(L_{\rm SN}/l_{\ast}) = L_{\rm SN}/l_{\ast}$, where
\begin{eqnarray}
\sigma^{2}_{\rm SN}(l) = V_{\rm vox}\phi_{\ast}\int_{0}^{l}dl^{\prime}l^{\prime \alpha+2}e^{-l^{\prime}}.
\label{sigma_sn}
\end{eqnarray}
$V_{\rm vox}$ is the comoving volume of a voxel, $\alpha$ is the faint-end slope parameter of the luminosity function. 
Observations can be divided into  four limiting regimes for optimal strategy, as shown in Table (\ref{regime}). 
In regime 1, the instrument noise is much smaller than $l_{\ast}$, and confusion effect is small, galaxy detection is optimal. 
In regime 2, the optimal strategy is somewhere intermediate between the intensity mapping and galaxy detection, 
because the voxels with $L \geqslant \sigma_{\rm L}$ will suffer from confusion noise. In regime 3, 
the instrument noise in a voxel is very large, $l_{\star} < \sigma_{\rm L}$, the intensity mapping will be the only choice. 
Regime 4 corresponds to a large effective number of galaxies per voxel, galaxy detection will suffer from large confusion noise, intensity mapping is optimal. 

\begin{figure}
  \centering
  \includegraphics[width=0.38\textwidth]{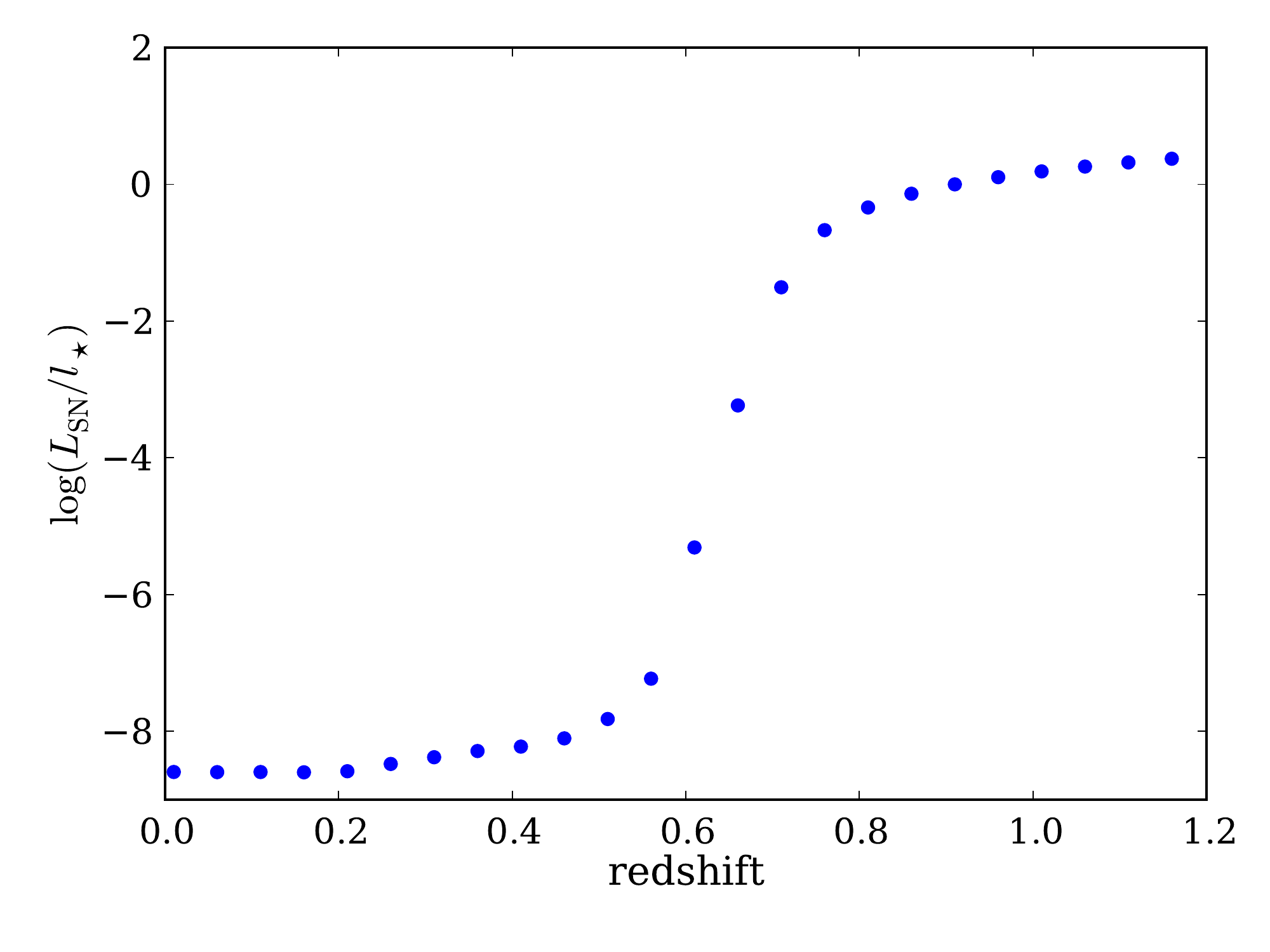}
  \includegraphics[width=0.38\textwidth]{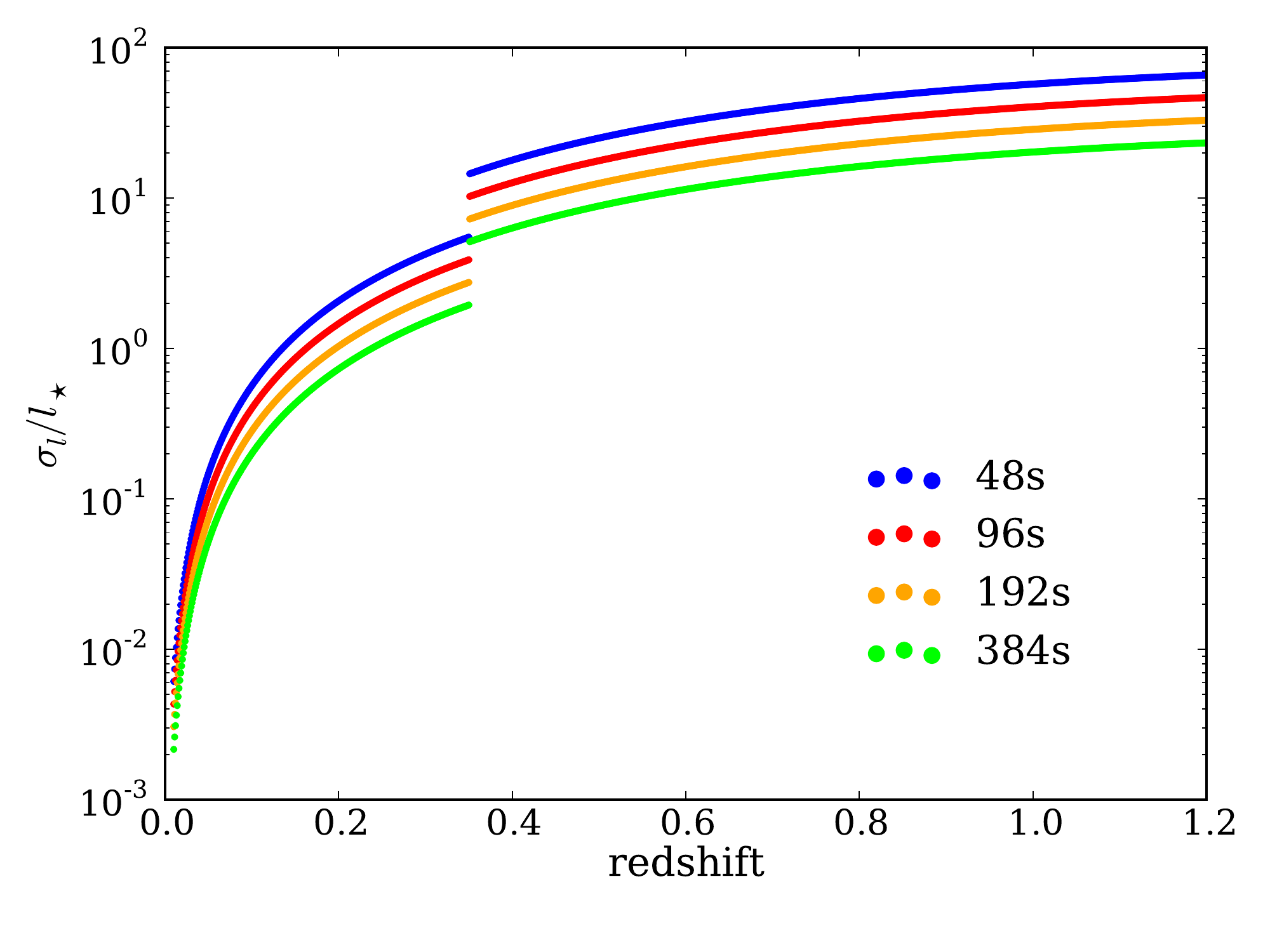}
  \caption{The redshift evolution of $L_{\rm SN}/l_{\ast}$ and $\sigma_{\rm L}/l_{\star}$, assuming an angular resolution of $0.08^\circ$ 
   and a bandwidth of 200km/s. The jump at redshift 0.35 in the bottom panel is due to the change of receiver system.}
  \label{lsn_sigmal_l}
\end{figure}

For the FAST HI survey, assuming a voxel of an angular resolution of $0.08^\circ$ and bandwidth of 200km/s, we compute the 
evolution of $L_{\rm SN}/l_{\ast}$ and $\sigma_{\rm L}/l_{\ast}$ in redshift and show the results in 
Figure. \ref{lsn_sigmal_l}. We use the luminosity function given by \citet{2003ApJ...592..819B}, 
with the following parameters: 
$\phi^{\ast} = 5.11\times10^{-3}h_{70}^3$ Mpc$^{-3}$, $\log (L_{\ast}/L_{\odot})=10.36+\log h_{70}$ and $\alpha= -1.05$. 
The redshift evolution of Schechter luminosity function parameters is modeled as\citep{1999ApJ...518..533L},
\begin{eqnarray}
     \alpha(z) &=& \alpha(z_{0}), \label{lf_evolution_alpha}\\
     M^{\star}(z) &=& M^{\star}(z_{0}) - Q (z - z_{0}), \label{lf_evolution_M}\\
     \varphi^{\star}(z) &=& \varphi^{\star}(z_{0})10^{0.4P (z - z_{0})}, \label{lf_evolution_phi}
\end{eqnarray}
where $P = 1.0$ and $Q = 1.03$ \citep{2015MNRAS.451.1540L} in the $r$-band. 

\begin{table}
 \caption{Four limiting regime defined by the relative value of the luminosity scale where the voxels are
  highly susceptible to shot noise, $L_{\rm SN}$, the rms noise per voxel, $\sigma_{\rm L}$ and the characteristic
   luminosity for a certain luminosity function, $l_{\star}\}$.}
 \label{regime}
 \begin{tabular}{ccc}
  \hline
  number&regime&optimal strategy\\
  \hline
  1&$L_{\rm SN} < \sigma_{\rm L} < l_{\star}$ & galaxy detection\\
  2&$\sigma_{\rm L} < L_{\rm SN} < l_{\star}$ & galaxy detection/intensity mapping$^a$\\
  3&$L_{\rm SN} < l_{\star} < \sigma_{\rm L}$ & intensity mapping\\
  4&$l_{\star} < L_{\rm SN}$ & intensity mapping\\
  \hline
  \multicolumn{3}{l}
  {\parbox{8cm}{$^a$ Here the optimal strategy is an intermediate between the  intensity mapping and galaxy detection observables.}}\\
 \end{tabular}
\end{table}

Figure. \ref{lsn_sigmal_l} shows the redshift evolution of $L_{\rm SN}/l_{\ast}$  (top panel), and $\sigma_{\rm L}/l_{\ast}$ (bottom panel).
We see $L_{\rm SN}\leqslant l_{\ast}$ at redshift $\lesssim$ 0.8. and $\sigma_{\rm L}\leqslant l_{\star}$ 
at redshift $\lesssim$ 0.13, 0.16, 0.19 and 0.23 for integration time of 48s, 96s, 192s and 384s respectively. 
Combining with Table (\ref{regime}), it shows for FAST the galaxy redshift survey is the optimal strategy at 
redshift $\lesssim$ 0.13, 0.16, 0.19 and 0.23 for survey with 48s, 96s, 192s and 384s integration time per beam. 
This is in agreement with what the projected error shown in Figure. \ref{projected_error_galaxy_count} and 
Figure. \ref{projected_error_im}.  We note that the voxel size in the projected error calculation for the 
intensity mapping is 2 $\Mpc/h$, which is larger than the voxel size for galaxy detection. This makes the redshift points 
 where intensity mapping is better than galaxy detection a little lower than the redshifts shown in Figure. \ref{lsn_sigmal_l}.

\subsection{Foreground}
One of the most challenging problems in intensity mapping experiment may be the contamination from the foreground radiation, which is 
several orders of magnitude larger in amplitude than the HI intensity signal.  It can in principle be subtracted, and the true signal recovered, 
based on the fact that the frequency dependency and some statistical properties of the foreground are different form the true signal. 
Sophisticated mathematical methods have been developed  \citep{2011PhRvD..83j3006L,2013MNRAS.434L..46S,2013MNRAS.429..165C,2014MNRAS.441.3271W,2014ApJ...781...57S,2015MNRAS.447..400A,2016ApJS..222....3Z}.
The intensity mapping experiment with the Green Bank Telescope (GBT) (\citet{2013MNRAS.434L..46S}) have shown that the 
 foregrounds can indeed be suppressed significantly,  though at present a positive detection of 21cm auto-power spectrum is yet to be achieved. 
Here we assume that after  a successful foreground subtraction, the contamination can be reduced to 
the thermal noise level \citep{2015MNRAS.454.3240B}.

\begin{figure}
  \centering
 \includegraphics[width=0.45\textwidth]{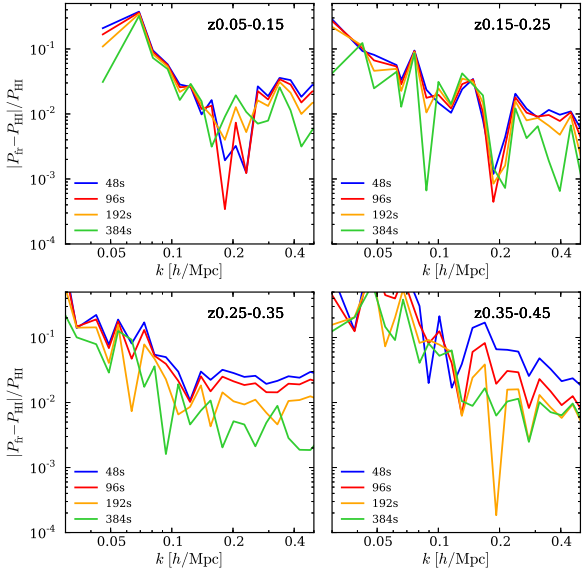}
   \caption{The relative error power spectra, defined by $(P_{\rm HI}-P_{\rm fr})/P_{\rm HI}$, where $P_{\rm HI}$ is the true 21cm power, 
  and $P_{\rm fr}$ is the foreground removed power  at different redshifts and integration time.}
  \label{spectra_with_foreground}
\end{figure}

To investigate the impact of foreground on the FAST IM survey, we made a simple 
test of foreground removal in our simulation. 
We produce the foreground with the global sky model (GSM) \citep{2008MNRAS.388..247D,2017MNRAS.464.3486Z}, 
convolved with a frequency-dependent beam and add to the noise-filled data cube. Then using a third order log-log
 polynomial fitting, we find the foreground can be removed effectively. The residue difference power spectra between 
 the original 21cm signal and the one obtained by removing the simulated foreground  with the polynomial fit 
 in the image cube are shown in Fig. \ref{spectra_with_foreground} with different integration times per beam for a Gaussian beam.  The power spectra show that the differences are small, at about 1 percent level at k $\approx$ 0.1 $h/\Mpc$ for these conditions. As expected, the amplitude of this error power spectra generally decreases  as integration time increases.
The differences at larger scales (smaller k) is larger, because a few large scale modes (modes with small k$_{\parallel}$) are removed by the foreground removal methods, 
in principle such large scale modes can be reconstructed via cosmic tidal reconstruction \citep{2018PhRvD..98d3511Z}. 
Using the foreground removed cube, we obtain the same constraints on the 
cosmological parameters in high accuracy.

\begin{figure}
  \centering
 \includegraphics[width=0.45\textwidth]{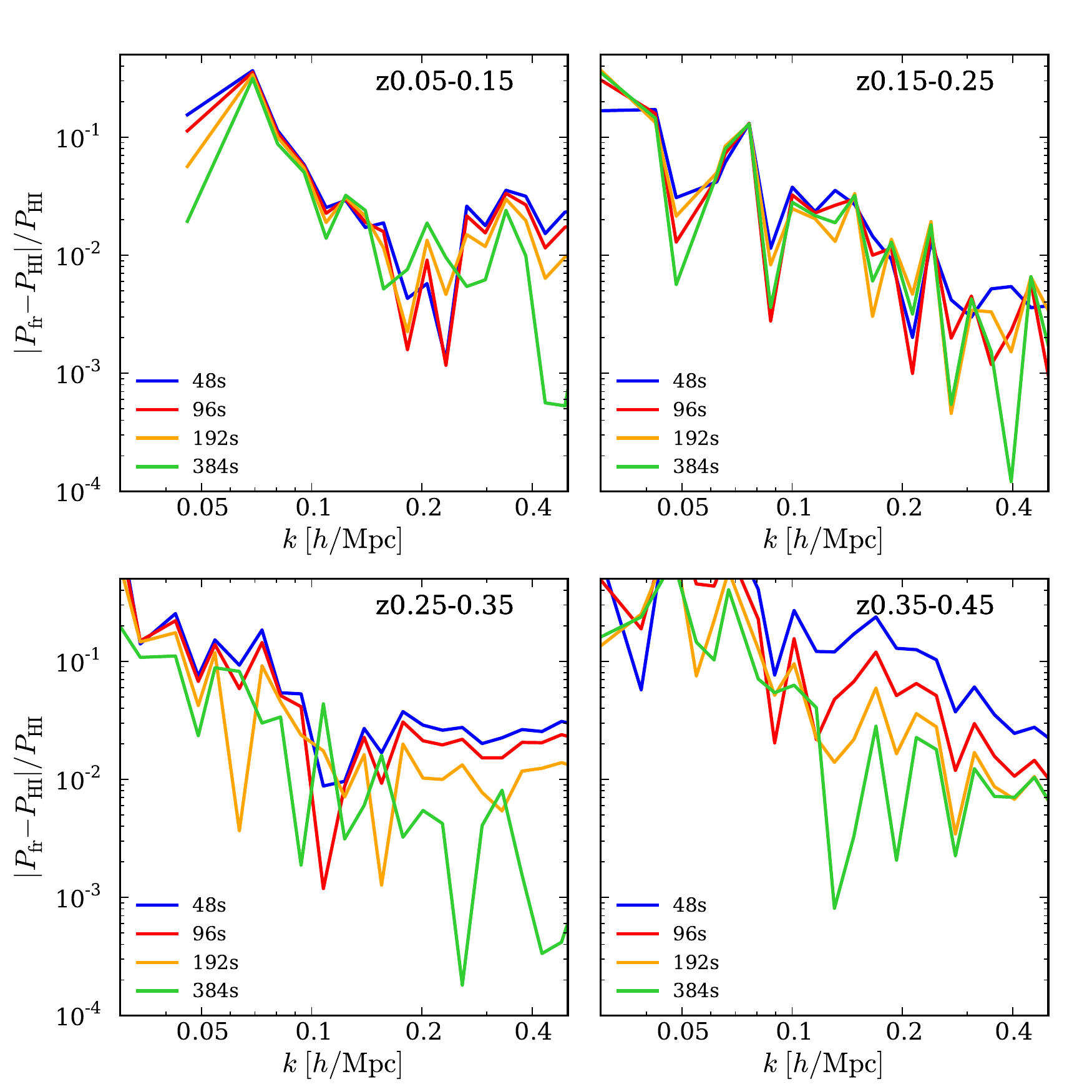}
   \caption{Same as Fig.\ref{spectra_with_foreground}, but the beam is modeled by Bessel function.}
  \label{spectra_with_foreground_bessel}
\end{figure}

Besides the Gaussian beam profile, we also made a test for the case of oscillatory side lobes. 
 In Fig.\ref{spectra_with_foreground_bessel} we make the same foreground subtraction excise, but for a beam function modeled
 using the Bessel function as $|2\times J_1(x)/x|^2$, where $x=3.23\times\theta/\sigma_b$. The result is qualitatively similar to the Gaussian case,
 though there are some differences.

The simulation of the foreground and its removal presented above may still
be too simplified. In reality, the the beam response
could be more complicated, and we only have an imperfect knowledge which must itself be determined from observation, 
and there are also irregularities in the beam and bandpass, polarization leakage, the $1/f$ noise and variations of the system gain, etc., making it much harder to remove. So the actual impact of the foreground could be higher than this simple estimate. 
 Obviously, the difficulty of the foreground removal depends on the design of the telescope. The stability, regularity and dynamic
range all affects the induced foreground contamination. A detailed study of foreground subtraction requires a realistic assessment of FAST telescope response,
which could only be obtained with actual observational data. This is beyond the scope of the present paper. However, the high sensitivity of the FAST is advantageous for 21cm extraction: the individual voxel signal-to-noise ratio is relatively high, so it 
is easier to be detected.

\section{Summary}
In this paper, we make a detailed study of large area drift scan HI survey with the FAST telescope. We considered using the existing L-band 19 beam 
receiver, the single feed wide band receiver for lower frequency (UHF), and also contemplated using a future UHF-band PAF receiver. 
We simulated observation of HI galaxies,  the number density of the detectable HI galaxies decreases rapidly as the redshift increases, and also 
due to the larger beam size and smaller galaxy size at higher redshifts, the mean number of galaxies within each voxel which we called confusion rate 
also increases. For the detected galaxies, there are also  measurement errors in both the direction of the line of sight and the direction perpendicular to it. 
We estimated such errors, but found that the main source of HI galaxy power spectrum error still comes from the shot noise on small scales and 
cosmic variance on large scales. We also considered the HI intensity mapping observation. 

The precision of the power spectrum measurement 
are forecasted using Fisher matrix for large survey areas, and we also make mock observations for both galaxy survey and intensity mapping survey
using simulation. With intensity mapping the power spectrum can be measured with high precision.  We find that the 
FAST can effectively detect the individual galaxy till z $\approx$  0.2, 0.25, 0.3 and 0.35 or map the large-scale structure 
with intensity mapping till z $\approx$ 0.35, 0.55, 0.75 and 1.05 respectively, if we assume 48s, 96s, 192s and 384s integration time per beam. 
Generally the HI intensity mapping observations can yield much more precise measurement, though 
the HI galaxy survey can also achieve nearly optimal measurement at lower redshifts,  with $z \lesssim 0.13, 0.16, 0.19$ and 0.23 for surveys 
with 48s, 96s, 192s and 384s integration time per L-band beam.  

We find that the FAST HI intensity mapping survey can produce a good measurement of the underlying power spectrum, and use the BAO method to 
measure the dark energy equation of state parameters. Such a measurement with a radio tracer is complementary to the optical BAO measurements, 
and reduce possible systematic errors. We also made a study of the impact of foregrounds on the measurement by simulation, and found that
it does not significantly affect the results, at least under the simplified conditions assumed in the study. 

With existing receivers the achievable precision is moderate, as it is largely limited to $z<0.35$, for higher redshifts 
the wide-band single feed would take too long time to complete the survey. However, if equipped with a UHF-band multi-beam receiver, higher redshifts 
can be observed more efficiently. We considered the case of a UHF-band PAF receiver with frequency coverage of $0.5 \sim 1.0\GHz$, and found that
the resulting survey may yield dark energy figure of merit up to $10^2$, comparable with the DETF stage IV results. 
This shows that a state-of-the-art PAF multi-beam receiver would be a very valuable addition to the FAST telescope.

\section*{Acknowledgements}
This work was first inspired by the late Prof. Rendong Nan. The parameters of the FAST receivers are provided to us by 
Di Li,  Youling Yue, Lei Qian and Jin Fan.  We also thank Shifan Zuo and Yichao Li for discussions.
This work is supported by the National Natural Science Foundation of China (NSFC) key
project grant 11633004, the Chinese Academy of Sciences (CAS) Frontier Science Key Project 
QYZDJ-SSW-SLH017 and the CAS Interdisciplinary Innovation Team grant (JCTD-2019-05), the NSFC-ISF joint research program No. 11761141012,  the MoST grant 2016YFE0100300, 
 the CAS Strategic Priority Research Program XDA15020200, the National Key R\&D Program 
2017YFA0402603, and the NSFC grant 11773034.

\bibliographystyle{mnras}
\bibliography{fast}

\begin{thebibliography}{}
\makeatletter
\relax
\def\mn@urlcharsother{\let\do\@makeother \do\$\do\&\do\#\do\^\do\_\do\%\do\~}
\def\mn@doi{\begingroup\mn@urlcharsother \@ifnextchar [ {\mn@doi@}
  {\mn@doi@[]}}
\def\mn@doi@[#1]#2{\def\@tempa{#1}\ifx\@tempa\@empty \href
  {http://dx.doi.org/#2} {doi:#2}\else \href {http://dx.doi.org/#2} {#1}\fi
  \endgroup}
\def\mn@eprint#1#2{\mn@eprint@#1:#2::\@nil}
\def\mn@eprint@arXiv#1{\href {http://arxiv.org/abs/#1} {{\tt arXiv:#1}}}
\def\mn@eprint@dblp#1{\href {http://dblp.uni-trier.de/rec/bibtex/#1.xml}
  {dblp:#1}}
\def\mn@eprint@#1:#2:#3:#4\@nil{\def\@tempa {#1}\def\@tempb {#2}\def\@tempc
  {#3}\ifx \@tempc \@empty \let \@tempc \@tempb \let \@tempb \@tempa \fi \ifx
  \@tempb \@empty \def\@tempb {arXiv}\fi \@ifundefined
  {mn@eprint@\@tempb}{\@tempb:\@tempc}{\expandafter \expandafter \csname
  mn@eprint@\@tempb\endcsname \expandafter{\@tempc}}}

\bibitem[\protect\citeauthoryear{{Alam} et~al.,}{{Alam}
  et~al.}{2017}]{2017MNRAS.470.2617A}
{Alam} S.,  et~al., 2017, \mn@doi [\mnras] {10.1093/mnras/stx721}, \href
  {http://adsabs.harvard.edu/abs/2017MNRAS.470.2617A} {470, 2617}

\bibitem[\protect\citeauthoryear{{Albrecht} et~al.,}{{Albrecht}
  et~al.}{2006}]{2006astro.ph..9591A}
{Albrecht} A.,  et~al., 2006, {Albrecht} A.,  et~al.,, \href
  {https://ui.adsabs.harvard.edu/abs/2006astro.ph..9591A} {pp
  astro--ph/0609591}

\bibitem[\protect\citeauthoryear{{Alonso}, {Bull}, {Ferreira}  \&
  {Santos}}{{Alonso} et~al.}{2015}]{2015MNRAS.447..400A}
{Alonso} D.,  {Bull} P.,  {Ferreira} P.~G.,   {Santos} M.~G.,  2015, \mn@doi
  [\mnras] {10.1093/mnras/stu2474}, \href
  {http://adsabs.harvard.edu/abs/2015MNRAS.447..400A} {447, 400}

\bibitem[\protect\citeauthoryear{{Ansari}, {Le Goff}, {Magneville}, {Moniez},
  {Palanque-Delabrouille}, {Rich}, {Ruhlmann-Kleider}  \& {Y{\`e}che}}{{Ansari}
  et~al.}{2008}]{2008arXiv0807.3614A}
{Ansari} R.,  {Le Goff} J.~.,  {Magneville} C.,  {Moniez} M.,
  {Palanque-Delabrouille} N.,  {Rich} J.,  {Ruhlmann-Kleider} V.,   {Y{\`e}che}
  C.,  2008, {Ansari} R.,  {Le Goff} J.~.,  {Magneville} C.,  {Moniez} M.,
  {Palanque-Delabrouille} N.,  {Rich} J.,  {Ruhlmann-Kleider} V.,   {Y{\`e}che}
  C., , \href {http://adsabs.harvard.edu/abs/2008arXiv0807.3614A} {}
  (\mn@eprint {arXiv} {0807.3614})

\bibitem[\protect\citeauthoryear{{Ansari} et~al.,}{{Ansari}
  et~al.}{2012}]{2012A&A...540A.129A}
{Ansari} R.,  et~al., 2012, \mn@doi [\aap] {10.1051/0004-6361/201117837}, \href
  {http://adsabs.harvard.edu/abs/2012A%26A...540A.129A} {540, A129}

\bibitem[\protect\citeauthoryear{{Ata} et~al.,}{{Ata}
  et~al.}{2018}]{2018MNRAS.473.4773A}
{Ata} M.,  et~al., 2018, \mn@doi [\mnras] {10.1093/mnras/stx2630}, \href
  {http://adsabs.harvard.edu/abs/2018MNRAS.473.4773A} {473, 4773}

\bibitem[\protect\citeauthoryear{{Bandura} et~al.,}{{Bandura}
  et~al.}{2014}]{2014SPIE.9145E..22B}
{Bandura} K.,  et~al., 2014, in Ground-based and Airborne Telescopes V. p.
  914522 (\mn@eprint {arXiv} {1406.2288}), \mn@doi{10.1117/12.2054950}

\bibitem[\protect\citeauthoryear{{Battye} et~al.,}{{Battye}
  et~al.}{2012}]{2012arXiv1209.1041B}
{Battye} R.~A.,  et~al., 2012.  (\mn@eprint {arXiv} {1209.1041})

\bibitem[\protect\citeauthoryear{{Battye} et~al.,}{{Battye}
  et~al.}{2016}]{2016arXiv161006826B}
{Battye} R.,  et~al., 2016.  (\mn@eprint {arXiv} {1610.06826})

\bibitem[\protect\citeauthoryear{{Beutler} et~al.,}{{Beutler}
  et~al.}{2011}]{2011MNRAS.416.3017B}
{Beutler} F.,  et~al., 2011, \mn@doi [\mnras]
  {10.1111/j.1365-2966.2011.19250.x}, \href
  {http://adsabs.harvard.edu/abs/2011MNRAS.416.3017B} {416, 3017}

\bibitem[\protect\citeauthoryear{{Bigot-Sazy} et~al.,}{{Bigot-Sazy}
  et~al.}{2015}]{2015MNRAS.454.3240B}
{Bigot-Sazy} M.-A.,  et~al., 2015, \mn@doi [\mnras] {10.1093/mnras/stv2153},
  \href {http://adsabs.harvard.edu/abs/2015MNRAS.454.3240B} {454, 3240}

\bibitem[\protect\citeauthoryear{{Blake} et~al.,}{{Blake}
  et~al.}{2011}]{2011MNRAS.418.1707B}
{Blake} C.,  et~al., 2011, \mn@doi [\mnras] {10.1111/j.1365-2966.2011.19592.x},
  \href {http://adsabs.harvard.edu/abs/2011MNRAS.418.1707B} {418, 1707}

\bibitem[\protect\citeauthoryear{{Blanton} et~al.,}{{Blanton}
  et~al.}{2003}]{2003ApJ...592..819B}
{Blanton} M.~R.,  et~al., 2003, \mn@doi [\apj] {10.1086/375776}, \href
  {http://adsabs.harvard.edu/abs/2003ApJ...592..819B} {592, 819}

\bibitem[\protect\citeauthoryear{{Blitz} \& {Rosolowsky}}{{Blitz} \&
  {Rosolowsky}}{2006}]{2006ApJ...650..933B}
{Blitz} L.,  {Rosolowsky} E.,  2006, \mn@doi [\apj] {10.1086/505417}, \href
  {http://adsabs.harvard.edu/abs/2006ApJ...650..933B} {650, 933}

\bibitem[\protect\citeauthoryear{{Braun}}{{Braun}}{2012}]{2012ApJ...749...87B}
{Braun} R.,  2012, \mn@doi [\apj] {10.1088/0004-637X/749/1/87}, \href
  {http://adsabs.harvard.edu/abs/2012ApJ...749...87B} {749, 87}

\bibitem[\protect\citeauthoryear{{Catinella}, {Giovanelli}  \&
  {Haynes}}{{Catinella} et~al.}{2006}]{2006ApJ...640..751C}
{Catinella} B.,  {Giovanelli} R.,   {Haynes} M.~P.,  2006, \mn@doi [\apj]
  {10.1086/500171}, \href {http://adsabs.harvard.edu/abs/2006ApJ...640..751C}
  {640, 751}

\bibitem[\protect\citeauthoryear{{Chang}, {Pen}, {Peterson}  \&
  {McDonald}}{{Chang} et~al.}{2008}]{2008PhRvL.100i1303C}
{Chang} T.-C.,  {Pen} U.-L.,  {Peterson} J.~B.,   {McDonald} P.,  2008, \mn@doi
  [Physical Review Letters] {10.1103/PhysRevLett.100.091303}, \href
  {http://adsabs.harvard.edu/abs/2008PhRvL.100i1303C} {100, 091303}

\bibitem[\protect\citeauthoryear{{Chapman} et~al.,}{{Chapman}
  et~al.}{2013}]{2013MNRAS.429..165C}
{Chapman} E.,  et~al., 2013, \mn@doi [\mnras] {10.1093/mnras/sts333}, \href
  {http://adsabs.harvard.edu/abs/2013MNRAS.429..165C} {429, 165}

\bibitem[\protect\citeauthoryear{{Chen}}{{Chen}}{2012}]{2012IJMPS..12..256C}
{Chen} X.,  2012, in International Journal of Modern Physics Conference Series.
  pp 256--263 (\mn@eprint {arXiv} {1212.6278}),
  \mn@doi{10.1142/S2010194512006459}

\bibitem[\protect\citeauthoryear{{Cheng}, {de Putter}, {Chang}  \&
  {Dore}}{{Cheng} et~al.}{2018}]{2018arXiv180906384C}
{Cheng} Y.-T.,  {de Putter} R.,  {Chang} T.-C.,   {Dore} O.,  2018, {Cheng}
  Y.-T.,  {de Putter} R.,  {Chang} T.-C.,   {Dore} O., , \href
  {http://adsabs.harvard.edu/abs/2018arXiv180906384C} {} (\mn@eprint {arXiv}
  {1809.06384})

\bibitem[\protect\citeauthoryear{{Chevallier} \& {Polarski}}{{Chevallier} \&
  {Polarski}}{2001}]{2001IJMPD..10..213C}
{Chevallier} M.,  {Polarski} D.,  2001, \mn@doi [International Journal of
  Modern Physics D] {10.1142/S0218271801000822}, \href
  {http://adsabs.harvard.edu/abs/2001IJMPD..10..213C} {10, 213}

\bibitem[\protect\citeauthoryear{{Cole} et~al.,}{{Cole}
  et~al.}{2005}]{2005MNRAS.362..505C}
{Cole} S.,  et~al., 2005, \mn@doi [\mnras] {10.1111/j.1365-2966.2005.09318.x},
  \href {http://adsabs.harvard.edu/abs/2005MNRAS.362..505C} {362, 505}

\bibitem[\protect\citeauthoryear{{Dav{\'e}}, {Rafieferantsoa}, {Thompson}  \&
  {Hopkins}}{{Dav{\'e}} et~al.}{2017}]{2017MNRAS.467..115D}
{Dav{\'e}} R.,  {Rafieferantsoa} M.~H.,  {Thompson} R.~J.,   {Hopkins} P.~F.,
  2017, \mn@doi [\mnras] {10.1093/mnras/stx108}, \href
  {http://adsabs.harvard.edu/abs/2017MNRAS.467..115D} {467, 115}

\bibitem[\protect\citeauthoryear{{Dawson} et~al.,}{{Dawson}
  et~al.}{2016}]{2016AJ....151...44D}
{Dawson} K.~S.,  et~al., 2016, \mn@doi [\aj] {10.3847/0004-6256/151/2/44},
  \href {http://adsabs.harvard.edu/abs/2016AJ....151...44D} {151, 44}

\bibitem[\protect\citeauthoryear{{De Lucia} \& {Blaizot}}{{De Lucia} \&
  {Blaizot}}{2007}]{2007MNRAS.375....2D}
{De Lucia} G.,  {Blaizot} J.,  2007, \mn@doi [\mnras]
  {10.1111/j.1365-2966.2006.11287.x}, \href
  {http://adsabs.harvard.edu/abs/2007MNRAS.375....2D} {375, 2}

\bibitem[\protect\citeauthoryear{{Delhaize}, {Meyer}, {Staveley-Smith}  \&
  {Boyle}}{{Delhaize} et~al.}{2013}]{2013MNRAS.433.1398D}
{Delhaize} J.,  {Meyer} M.~J.,  {Staveley-Smith} L.,   {Boyle} B.~J.,  2013,
  \mn@doi [\mnras] {10.1093/mnras/stt810}, \href
  {http://adsabs.harvard.edu/abs/2013MNRAS.433.1398D} {433, 1398}

\bibitem[\protect\citeauthoryear{{Drinkwater} et~al.,}{{Drinkwater}
  et~al.}{2010}]{2010MNRAS.401.1429D}
{Drinkwater} M.~J.,  et~al., 2010, \mn@doi [\mnras]
  {10.1111/j.1365-2966.2009.15754.x}, \href
  {http://adsabs.harvard.edu/abs/2010MNRAS.401.1429D} {401, 1429}

\bibitem[\protect\citeauthoryear{{Duffy}, {Battye}, {Davies}, {Moss}  \&
  {Wilkinson}}{{Duffy} et~al.}{2008}]{2008MNRAS.383..150D}
{Duffy} A.~R.,  {Battye} R.~A.,  {Davies} R.~D.,  {Moss} A.,   {Wilkinson}
  P.~N.,  2008, \mn@doi [\mnras] {10.1111/j.1365-2966.2007.12537.x}, \href
  {http://adsabs.harvard.edu/abs/2008MNRAS.383..150D} {383, 150}

\bibitem[\protect\citeauthoryear{{Duffy}, {Meyer}, {Staveley-Smith}, {Bernyk},
  {Croton}, {Koribalski}, {Gerstmann}  \& {Westerlund}}{{Duffy}
  et~al.}{2012}]{2012MNRAS.426.3385D}
{Duffy} A.~R.,  {Meyer} M.~J.,  {Staveley-Smith} L.,  {Bernyk} M.,  {Croton}
  D.~J.,  {Koribalski} B.~S.,  {Gerstmann} D.,   {Westerlund} S.,  2012,
  \mn@doi [\mnras] {10.1111/j.1365-2966.2012.21987.x}, \href
  {http://adsabs.harvard.edu/abs/2012MNRAS.426.3385D} {426, 3385}

\bibitem[\protect\citeauthoryear{{Eisenstein} et~al.,}{{Eisenstein}
  et~al.}{2005}]{2005ApJ...633..560E}
{Eisenstein} D.~J.,  et~al., 2005, \mn@doi [\apj] {10.1086/466512}, \href
  {http://adsabs.harvard.edu/abs/2005ApJ...633..560E} {633, 560}

\bibitem[\protect\citeauthoryear{{Elson}, {Blyth}  \& {Baker}}{{Elson}
  et~al.}{2016}]{2016MNRAS.460.4366E}
{Elson} E.~C.,  {Blyth} S.~L.,   {Baker} A.~J.,  2016, \mn@doi [\mnras]
  {10.1093/mnras/stw1291}, \href
  {http://adsabs.harvard.edu/abs/2016MNRAS.460.4366E} {460, 4366}

\bibitem[\protect\citeauthoryear{{Feldman}, {Kaiser}  \& {Peacock}}{{Feldman}
  et~al.}{1994}]{1994ApJ..426...23F}
{Feldman} H.~A.,  {Kaiser} N.,   {Peacock} J.~A.,  1994, \mn@doi [\apj]
  {10.1086/174036}, \href {http://adsabs.harvard.edu/abs/1994ApJ...426...23F}
  {426, 23}

\bibitem[\protect\citeauthoryear{{Freudling} et~al.,}{{Freudling}
  et~al.}{2011}]{2011ApJ...727...40F}
{Freudling} W.,  et~al., 2011, \mn@doi [\apj] {10.1088/0004-637X/727/1/40},
  \href {http://adsabs.harvard.edu/abs/2011ApJ...727...40F} {727, 40}

\bibitem[\protect\citeauthoryear{{Giovanelli} \& {Haynes}}{{Giovanelli} \&
  {Haynes}}{2002}]{2002ApJ...571L.107G}
{Giovanelli} R.,  {Haynes} M.~P.,  2002, \mn@doi [\apjl] {10.1086/341368},
  \href {http://adsabs.harvard.edu/abs/2002ApJ...571L.107G} {571, L107}

\bibitem[\protect\citeauthoryear{{Giovanelli} et~al.,}{{Giovanelli}
  et~al.}{2005}]{2005AJ....130.2598G}
{Giovanelli} R.,  et~al., 2005, \mn@doi [\aj] {10.1086/497431}, \href
  {http://adsabs.harvard.edu/abs/2005AJ....130.2598G} {130, 2598}

\bibitem[\protect\citeauthoryear{{Giovanelli} et~al.,}{{Giovanelli}
  et~al.}{2007}]{2007AJ....133.2569G}
{Giovanelli} R.,  et~al., 2007, \mn@doi [\aj] {10.1086/516635}, \href
  {http://adsabs.harvard.edu/abs/2007AJ....133.2569G} {133, 2569}

\bibitem[\protect\citeauthoryear{{Glazebrook} et~al.,}{{Glazebrook}
  et~al.}{2007}]{2007ASPC..379...72G}
{Glazebrook} K.,  et~al., 2007, in {Metcalfe} N.,  {Shanks} T.,  eds,
  Astronomical Society of the Pacific Conference Series Vol. 379, Cosmic
  Frontiers. p.~72 (\mn@eprint {} {astro-ph/0701876})

\bibitem[\protect\citeauthoryear{{Goldberg} \& {Strauss}}{{Goldberg} \&
  {Strauss}}{1998}]{1998ApJ...495...29G}
{Goldberg} D.~M.,  {Strauss} M.~A.,  1998, \mn@doi [\apj] {10.1086/305284},
  \href {http://adsabs.harvard.edu/abs/1998ApJ...495...29G} {495, 29}

\bibitem[\protect\citeauthoryear{{Gunn} \& {Weinberg}}{{Gunn} \&
  {Weinberg}}{1995}]{1995wfsd.conf....3G}
{Gunn} J.,  {Weinberg} D.,  1995, in {Maddox} S.~J.,  {Aragon-Salamanca} A.,
  eds, Wide Field Spectroscopy and the Distant Universe. p.~3 (\mn@eprint {}
  {astro-ph/9412080})

\bibitem[\protect\citeauthoryear{{Hoppmann}, {Staveley-Smith}, {Freudling},
  {Zwaan}, {Minchin}  \& {Calabretta}}{{Hoppmann}
  et~al.}{2015}]{2015MNRAS.452.3726H}
{Hoppmann} L.,  {Staveley-Smith} L.,  {Freudling} W.,  {Zwaan} M.~A.,
  {Minchin} R.~F.,   {Calabretta} M.~R.,  2015, \mn@doi [\mnras]
  {10.1093/mnras/stv1084}, \href
  {http://adsabs.harvard.edu/abs/2015MNRAS.452.3726H} {452, 3726}

\bibitem[\protect\citeauthoryear{{Hu} et~al.,}{{Hu}
  et~al.}{2019}]{2019MNRAS.489.1619H}
{Hu} W.,  et~al., 2019, \mn@doi [\mnras] {10.1093/mnras/stz2038}, \href
  {https://ui.adsabs.harvard.edu/abs/2019MNRAS.489.1619H} {489, 1619}

\bibitem[\protect\citeauthoryear{{Jarvis} et~al.,}{{Jarvis}
  et~al.}{2014}]{2014arXiv1401.4018J}
{Jarvis} M.~J.,  et~al., 2014, {Jarvis} M.~J.,  et~al.,, \href
  {http://adsabs.harvard.edu/abs/2014arXiv1401.4018J} {} (\mn@eprint {arXiv}
  {1401.4018})

\bibitem[\protect\citeauthoryear{{Jones}, {Haynes}, {Giovanelli}  \&
  {Moorman}}{{Jones} et~al.}{2018}]{2018MNRAS.tmp..502J}
{Jones} M.~G.,  {Haynes} M.~P.,  {Giovanelli} R.,   {Moorman} C.,  2018,
  \mn@doi [\mnras] {10.1093/mnras/sty521}, \href
  {https://ui.adsabs.harvard.edu/abs/2018MNRAS.477....2J} {477, 2}

\bibitem[\protect\citeauthoryear{{Kanekar}, {Sethi}  \&
  {Dwarakanath}}{{Kanekar} et~al.}{2016}]{2016ApJ...818L..28K}
{Kanekar} N.,  {Sethi} S.,   {Dwarakanath} K.~S.,  2016, \mn@doi [\apjl]
  {10.3847/2041-8205/818/2/L28}, \href
  {http://adsabs.harvard.edu/abs/2016ApJ...818L..28K} {818, L28}

\bibitem[\protect\citeauthoryear{{Lah} et~al.,}{{Lah}
  et~al.}{2007}]{2007MNRAS.376.1357L}
{Lah} P.,  et~al., 2007, \mn@doi [\mnras] {10.1111/j.1365-2966.2007.11540.x},
  \href {http://adsabs.harvard.edu/abs/2007MNRAS.376.1357L} {376, 1357}

\bibitem[\protect\citeauthoryear{{Leroy}, {Walter}, {Brinks}, {Bigiel}, {de
  Blok}, {Madore}  \& {Thornley}}{{Leroy} et~al.}{2008}]{2008AJ....136.2782L}
{Leroy} A.~K.,  {Walter} F.,  {Brinks} E.,  {Bigiel} F.,  {de Blok} W.~J.~G.,
  {Madore} B.,   {Thornley} M.~D.,  2008, \mn@doi [\aj]
  {10.1088/0004-6256/136/6/2782}, \href
  {http://adsabs.harvard.edu/abs/2008AJ....136.2782L} {136, 2782}

\bibitem[\protect\citeauthoryear{{Li} et~al.,}{{Li}
  et~al.}{2018}]{2018IMMag..19..112L}
{Li} D.,  et~al., 2018, \mn@doi [IEEE Microwave Magazine]
  {10.1109/MMM.2018.2802178}, \href
  {https://ui.adsabs.harvard.edu/abs/2018IMMag..19..112L} {19, 112}

\bibitem[\protect\citeauthoryear{{Lin}, {Yee}, {Carlberg}, {Morris}, {Sawicki},
  {Patton}, {Wirth}  \& {Shepherd}}{{Lin} et~al.}{1999}]{1999ApJ...518..533L}
{Lin} H.,  {Yee} H.~K.~C.,  {Carlberg} R.~G.,  {Morris} S.~L.,  {Sawicki} M.,
  {Patton} D.~R.,  {Wirth} G.,   {Shepherd} C.~W.,  1999, \mn@doi [\apj]
  {10.1086/307297}, \href {http://adsabs.harvard.edu/abs/1999ApJ...518..533L}
  {518, 533}

\bibitem[\protect\citeauthoryear{{Liu} \& {Tegmark}}{{Liu} \&
  {Tegmark}}{2011}]{2011PhRvD..83j3006L}
{Liu} A.,  {Tegmark} M.,  2011, \mn@doi [\prd] {10.1103/PhysRevD.83.103006},
  \href {http://adsabs.harvard.edu/abs/2011PhRvD..83j3006L} {83, 103006}

\bibitem[\protect\citeauthoryear{{Loveday} et~al.,}{{Loveday}
  et~al.}{2015}]{2015MNRAS.451.1540L}
{Loveday} J.,  et~al., 2015, \mn@doi [\mnras] {10.1093/mnras/stv1013}, \href
  {http://adsabs.harvard.edu/abs/2015MNRAS.451.1540L} {451, 1540}

\bibitem[\protect\citeauthoryear{{Martin}, {Papastergis}, {Giovanelli},
  {Haynes}, {Springob}  \& {Stierwalt}}{{Martin}
  et~al.}{2010}]{2010ApJ...723.1359M}
{Martin} A.~M.,  {Papastergis} E.,  {Giovanelli} R.,  {Haynes} M.~P.,
  {Springob} C.~M.,   {Stierwalt} S.,  2010, \mn@doi [\apj]
  {10.1088/0004-637X/723/2/1359}, \href
  {http://adsabs.harvard.edu/abs/2010ApJ...723.1359M} {723, 1359}

\bibitem[\protect\citeauthoryear{{Meiksin}, {White}  \& {Peacock}}{{Meiksin}
  et~al.}{1999}]{1999MNRAS.304..851M}
{Meiksin} A.,  {White} M.,   {Peacock} J.~A.,  1999, \mn@doi [\mnras]
  {10.1046/j.1365-8711.1999.02369.x}, \href
  {http://adsabs.harvard.edu/abs/1999MNRAS.304..851M} {304, 851}

\bibitem[\protect\citeauthoryear{{Meyer} et~al.,}{{Meyer}
  et~al.}{2004}]{2004MNRAS.350.1195M}
{Meyer} M.~J.,  et~al., 2004, \mn@doi [\mnras]
  {10.1111/j.1365-2966.2004.07710.x}, \href
  {http://adsabs.harvard.edu/abs/2004MNRAS.350.1195M} {350, 1195}

\bibitem[\protect\citeauthoryear{{Mo} \& {White}}{{Mo} \&
  {White}}{2002}]{2002MNRAS.336..112M}
{Mo} H.~J.,  {White} S.~D.~M.,  2002, \mn@doi [\mnras]
  {10.1046/j.1365-8711.2002.05723.x}, \href
  {http://adsabs.harvard.edu/abs/2002MNRAS.336..112M} {336, 112}

\bibitem[\protect\citeauthoryear{{More}, {van den Bosch}, {Cacciato}, {More},
  {Mo}  \& {Yang}}{{More} et~al.}{2013}]{2013MNRAS.430..747M}
{More} S.,  {van den Bosch} F.~C.,  {Cacciato} M.,  {More} A.,  {Mo} H.,
  {Yang} X.,  2013, \mn@doi [\mnras] {10.1093/mnras/sts697}, \href
  {http://adsabs.harvard.edu/abs/2013MNRAS.430..747M} {430, 747}

\bibitem[\protect\citeauthoryear{{Nan} et~al.,}{{Nan}
  et~al.}{2011}]{2011IJMPD..20..989N}
{Nan} R.,  et~al., 2011, \mn@doi [International Journal of Modern Physics D]
  {10.1142/S0218271811019335}, \href
  {http://adsabs.harvard.edu/abs/2011IJMPD..20..989N} {20, 989}

\bibitem[\protect\citeauthoryear{{Neeleman}, {Prochaska}, {Ribaudo}, {Lehner},
  {Howk}, {Rafelski}  \& {Kanekar}}{{Neeleman}
  et~al.}{2016}]{2016ApJ...818..113N}
{Neeleman} M.,  {Prochaska} J.~X.,  {Ribaudo} J.,  {Lehner} N.,  {Howk} J.~C.,
  {Rafelski} M.,   {Kanekar} N.,  2016, \mn@doi [\apj]
  {10.3847/0004-637X/818/2/113}, \href
  {http://adsabs.harvard.edu/abs/2016ApJ...818..113N} {818, 113}

\bibitem[\protect\citeauthoryear{{Newburgh} et~al.,}{{Newburgh}
  et~al.}{2016}]{2016SPIE.9906E..5XN}
{Newburgh} L.~B.,  et~al., 2016, in Ground-based and Airborne Telescopes VI. p.
  99065X (\mn@eprint {arXiv} {1607.02059}), \mn@doi{10.1117/12.2234286}

\bibitem[\protect\citeauthoryear{{Nyland} et~al.,}{{Nyland}
  et~al.}{2017}]{2017MNRAS.464.1029N}
{Nyland} K.,  et~al., 2017, \mn@doi [\mnras] {10.1093/mnras/stw2385}, \href
  {http://adsabs.harvard.edu/abs/2017MNRAS.464.1029N} {464, 1029}

\bibitem[\protect\citeauthoryear{{Obreschkow} \& {Meyer}}{{Obreschkow} \&
  {Meyer}}{2014}]{2014arXiv1406.0966O}
{Obreschkow} D.,  {Meyer} M.,  2014, {Obreschkow} D.,  {Meyer} M., , \href
  {http://adsabs.harvard.edu/abs/2014arXiv1406.0966O} {} (\mn@eprint {arXiv}
  {1406.0966})

\bibitem[\protect\citeauthoryear{{Obreschkow}, {Croton}, {De Lucia}, {Khochfar}
   \& {Rawlings}}{{Obreschkow} et~al.}{2009a}]{2009ApJ...698.1467O}
{Obreschkow} D.,  {Croton} D.,  {De Lucia} G.,  {Khochfar} S.,   {Rawlings} S.,
   2009a, \mn@doi [\apj] {10.1088/0004-637X/698/2/1467}, \href
  {http://adsabs.harvard.edu/abs/2009ApJ...698.1467O} {698, 1467}

\bibitem[\protect\citeauthoryear{{Obreschkow}, {Heywood}, {Kl{\"o}ckner}  \&
  {Rawlings}}{{Obreschkow} et~al.}{2009b}]{2009ApJ...702.1321O}
{Obreschkow} D.,  {Heywood} I.,  {Kl{\"o}ckner} H.-R.,   {Rawlings} S.,  2009b,
  \mn@doi [\apj] {10.1088/0004-637X/702/2/1321}, \href
  {http://adsabs.harvard.edu/abs/2009ApJ...702.1321O} {702, 1321}

\bibitem[\protect\citeauthoryear{{Obreschkow}, {Kl{\"o}ckner}, {Heywood},
  {Levrier}  \& {Rawlings}}{{Obreschkow} et~al.}{2009c}]{2009ApJ...703.1890O}
{Obreschkow} D.,  {Kl{\"o}ckner} H.-R.,  {Heywood} I.,  {Levrier} F.,
  {Rawlings} S.,  2009c, \mn@doi [\apj] {10.1088/0004-637X/703/2/1890}, \href
  {http://adsabs.harvard.edu/abs/2009ApJ...703.1890O} {703, 1890}

\bibitem[\protect\citeauthoryear{{Percival} et~al.,}{{Percival}
  et~al.}{2001}]{2001MNRAS.327.1297P}
{Percival} W.~J.,  et~al., 2001, \mn@doi [\mnras]
  {10.1046/j.1365-8711.2001.04827.x}, \href
  {http://adsabs.harvard.edu/abs/2001MNRAS.327.1297P} {327, 1297}

\bibitem[\protect\citeauthoryear{{Percival} et~al.,}{{Percival}
  et~al.}{2010}]{2010MNRAS.401.2148P}
{Percival} W.~J.,  et~al., 2010, \mn@doi [\mnras]
  {10.1111/j.1365-2966.2009.15812.x}, \href
  {http://adsabs.harvard.edu/abs/2010MNRAS.401.2148P} {401, 2148}

\bibitem[\protect\citeauthoryear{{Rao}, {Turnshek}  \& {Nestor}}{{Rao}
  et~al.}{2006}]{2006ApJ...636..610R}
{Rao} S.~M.,  {Turnshek} D.~A.,   {Nestor} D.~B.,  2006, \mn@doi [\apj]
  {10.1086/498132}, \href {http://adsabs.harvard.edu/abs/2006ApJ...636..610R}
  {636, 610}

\bibitem[\protect\citeauthoryear{{Rao}, {Turnshek}, {Sardane}  \&
  {Monier}}{{Rao} et~al.}{2017}]{2017MNRAS.471.3428R}
{Rao} S.~M.,  {Turnshek} D.~A.,  {Sardane} G.~M.,   {Monier} E.~M.,  2017,
  \mn@doi [\mnras] {10.1093/mnras/stx1787}, \href
  {http://adsabs.harvard.edu/abs/2017MNRAS.471.3428R} {471, 3428}

\bibitem[\protect\citeauthoryear{{Rhee}, {Zwaan}, {Briggs}, {Chengalur}, {Lah},
  {Oosterloo}  \& {van der Hulst}}{{Rhee} et~al.}{2013}]{2013MNRAS.435.2693R}
{Rhee} J.,  {Zwaan} M.~A.,  {Briggs} F.~H.,  {Chengalur} J.~N.,  {Lah} P.,
  {Oosterloo} T.,   {van der Hulst} T.,  2013, \mn@doi [\mnras]
  {10.1093/mnras/stt1481}, \href
  {http://adsabs.harvard.edu/abs/2013MNRAS.435.2693R} {435, 2693}

\bibitem[\protect\citeauthoryear{{Rhee}, {Lah}, {Chengalur}, {Briggs}  \&
  {Colless}}{{Rhee} et~al.}{2016}]{2016MNRAS.460.2675R}
{Rhee} J.,  {Lah} P.,  {Chengalur} J.~N.,  {Briggs} F.~H.,   {Colless} M.,
  2016, \mn@doi [\mnras] {10.1093/mnras/stw1097}, \href
  {http://adsabs.harvard.edu/abs/2016MNRAS.460.2675R} {460, 2675}

\bibitem[\protect\citeauthoryear{{Rhee}, {Lah}, {Briggs}, {Chengalur},
  {Colless}, {Willner}, {Ashby}  \& {Le F{\`e}vre}}{{Rhee}
  et~al.}{2018}]{2018MNRAS.473.1879R}
{Rhee} J.,  {Lah} P.,  {Briggs} F.~H.,  {Chengalur} J.~N.,  {Colless} M.,
  {Willner} S.~P.,  {Ashby} M.~L.~N.,   {Le F{\`e}vre} O.,  2018, \mn@doi
  [\mnras] {10.1093/mnras/stx2461}, \href
  {http://adsabs.harvard.edu/abs/2018MNRAS.473.1879R} {473, 1879}

\bibitem[\protect\citeauthoryear{{Ross}, {Samushia}, {Howlett}, {Percival},
  {Burden}  \& {Manera}}{{Ross} et~al.}{2015}]{2015MNRAS.449..835R}
{Ross} A.~J.,  {Samushia} L.,  {Howlett} C.,  {Percival} W.~J.,  {Burden} A.,
  {Manera} M.,  2015, \mn@doi [\mnras] {10.1093/mnras/stv154}, \href
  {http://adsabs.harvard.edu/abs/2015MNRAS.449..835R} {449, 835}

\bibitem[\protect\citeauthoryear{{Saintonge}}{{Saintonge}}{2007}]{2007AJ....133.2087S}
{Saintonge} A.,  2007, \mn@doi [\aj] {10.1086/513515}, \href
  {http://adsabs.harvard.edu/abs/2007AJ....133.2087S} {133, 2087}

\bibitem[\protect\citeauthoryear{{Santos} et~al.,}{{Santos}
  et~al.}{2015}]{2015aska.confE..19S}
{Santos} M.,  et~al., 2015. p.~19 (\mn@eprint {arXiv} {1501.03989})

\bibitem[\protect\citeauthoryear{{Schlegel}, {White}  \&
  {Eisenstein}}{{Schlegel} et~al.}{2009}]{2009astro2010S.314S}
{Schlegel} D.,  {White} M.,   {Eisenstein} D.,  2009, in astro2010: The
  Astronomy and Astrophysics Decadal Survey.  (\mn@eprint {arXiv} {0902.4680})

\bibitem[\protect\citeauthoryear{{Seo} \& {Eisenstein}}{{Seo} \&
  {Eisenstein}}{2003}]{2003ApJ...598..720S}
{Seo} H.-J.,  {Eisenstein} D.~J.,  2003, \mn@doi [\apj] {10.1086/379122}, \href
  {http://adsabs.harvard.edu/abs/2003ApJ...598..720S} {598, 720}

\bibitem[\protect\citeauthoryear{{Seo}, {Dodelson}, {Marriner}, {Mcginnis},
  {Stebbins}, {Stoughton}  \& {Vallinotto}}{{Seo}
  et~al.}{2010}]{2010ApJ...721..164S}
{Seo} H.-J.,  {Dodelson} S.,  {Marriner} J.,  {Mcginnis} D.,  {Stebbins} A.,
  {Stoughton} C.,   {Vallinotto} A.,  2010, \mn@doi [\apj]
  {10.1088/0004-637X/721/1/164}, \href
  {http://adsabs.harvard.edu/abs/2010ApJ...721..164S} {721, 164}

\bibitem[\protect\citeauthoryear{{Serra} et~al.,}{{Serra}
  et~al.}{2012}]{2012MNRAS.422.1835S}
{Serra} P.,  et~al., 2012, \mn@doi [\mnras] {10.1111/j.1365-2966.2012.20219.x},
  \href {http://adsabs.harvard.edu/abs/2012MNRAS.422.1835S} {422, 1835}

\bibitem[\protect\citeauthoryear{{Shaw}, {Sigurdson}, {Pen}, {Stebbins}  \&
  {Sitwell}}{{Shaw} et~al.}{2014}]{2014ApJ...781...57S}
{Shaw} J.~R.,  {Sigurdson} K.,  {Pen} U.-L.,  {Stebbins} A.,   {Sitwell} M.,
  2014, \mn@doi [\apj] {10.1088/0004-637X/781/2/57}, \href
  {http://adsabs.harvard.edu/abs/2014ApJ...781...57S} {781, 57}

\bibitem[\protect\citeauthoryear{{Sheth} \& {Tormen}}{{Sheth} \&
  {Tormen}}{2002}]{2002MNRAS.329...61S}
{Sheth} R.~K.,  {Tormen} G.,  2002, \mn@doi [\mnras]
  {10.1046/j.1365-8711.2002.04950.x}, \href
  {http://adsabs.harvard.edu/abs/2002MNRAS.329...61S} {329, 61}

\bibitem[\protect\citeauthoryear{{Smoot} \& {Debono}}{{Smoot} \&
  {Debono}}{2017}]{2017A&A...597A.136S}
{Smoot} G.~F.,  {Debono} I.,  2017, \mn@doi [\aap]
  {10.1051/0004-6361/201526794}, \href
  {http://adsabs.harvard.edu/abs/2017A%26A...597A.136S} {597, A136}

\bibitem[\protect\citeauthoryear{{Springel} et~al.,}{{Springel}
  et~al.}{2005}]{2005Natur.435..629S}
{Springel} V.,  et~al., 2005, \mn@doi [\nat] {10.1038/nature03597}, \href
  {http://adsabs.harvard.edu/abs/2005Natur.435..629S} {435, 629}

\bibitem[\protect\citeauthoryear{{Switzer} et~al.,}{{Switzer}
  et~al.}{2013}]{2013MNRAS.434L..46S}
{Switzer} E.~R.,  et~al., 2013, \mn@doi [\mnras] {10.1093/mnrasl/slt074}, \href
  {http://adsabs.harvard.edu/abs/2013MNRAS.434L..46S} {434, L46}

\bibitem[\protect\citeauthoryear{{Tacconi} et~al.,}{{Tacconi}
  et~al.}{2006}]{2006ApJ...640..228T}
{Tacconi} L.~J.,  et~al., 2006, \mn@doi [\apj] {10.1086/499933}, \href
  {http://adsabs.harvard.edu/abs/2006ApJ...640..228T} {640, 228}

\bibitem[\protect\citeauthoryear{{Tegmark}}{{Tegmark}}{1997}]{1997PhRvL..79.3806T}
{Tegmark} M.,  1997, \mn@doi [Physical Review Letters]
  {10.1103/PhysRevLett.79.3806}, \href
  {http://adsabs.harvard.edu/abs/1997PhRvL..79.3806T} {79, 3806}

\bibitem[\protect\citeauthoryear{{Tegmark} et~al.,}{{Tegmark}
  et~al.}{2004}]{2004ApJ...606..702T}
{Tegmark} M.,  et~al., 2004, \mn@doi [\apj] {10.1086/382125}, \href
  {http://adsabs.harvard.edu/abs/2004ApJ...606..702T} {606, 702}

\bibitem[\protect\citeauthoryear{{Tinker}, {Kravtsov}, {Klypin}, {Abazajian},
  {Warren}, {Yepes}, {Gottl{\"o}ber}  \& {Holz}}{{Tinker}
  et~al.}{2008}]{2008ApJ...688..709T}
{Tinker} J.,  {Kravtsov} A.~V.,  {Klypin} A.,  {Abazajian} K.,  {Warren} M.,
  {Yepes} G.,  {Gottl{\"o}ber} S.,   {Holz} D.~E.,  2008, \mn@doi [\apj]
  {10.1086/591439}, \href {http://adsabs.harvard.edu/abs/2008ApJ...688..709T}
  {688, 709}

\bibitem[\protect\citeauthoryear{{Tinker}, {Robertson}, {Kravtsov}, {Klypin},
  {Warren}, {Yepes}  \& {Gottl{\"o}ber}}{{Tinker}
  et~al.}{2010}]{2010ApJ...724..878T}
{Tinker} J.~L.,  {Robertson} B.~E.,  {Kravtsov} A.~V.,  {Klypin} A.,  {Warren}
  M.~S.,  {Yepes} G.,   {Gottl{\"o}ber} S.,  2010, \mn@doi [\apj]
  {10.1088/0004-637X/724/2/878}, \href
  {http://adsabs.harvard.edu/abs/2010ApJ...724..878T} {724, 878}

\bibitem[\protect\citeauthoryear{{Westmeier}, {Jurek}, {Obreschkow},
  {Koribalski}  \& {Staveley-Smith}}{{Westmeier}
  et~al.}{2014}]{2014MNRAS.438.1176W}
{Westmeier} T.,  {Jurek} R.,  {Obreschkow} D.,  {Koribalski} B.~S.,
  {Staveley-Smith} L.,  2014, \mn@doi [\mnras] {10.1093/mnras/stt2266}, \href
  {http://adsabs.harvard.edu/abs/2014MNRAS.438.1176W} {438, 1176}

\bibitem[\protect\citeauthoryear{{Wolz}, {Abdalla}, {Blake}, {Shaw}, {Chapman}
  \& {Rawlings}}{{Wolz} et~al.}{2014}]{2014MNRAS.441.3271W}
{Wolz} L.,  {Abdalla} F.~B.,  {Blake} C.,  {Shaw} J.~R.,  {Chapman} E.,
  {Rawlings} S.,  2014, \mn@doi [\mnras] {10.1093/mnras/stu792}, \href
  {http://adsabs.harvard.edu/abs/2014MNRAS.441.3271W} {441, 3271}

\bibitem[\protect\citeauthoryear{Wu, Jin, Fan, Zhao, Yu  \& Du}{Wu
  et~al.}{2016}]{wu2016}
Wu Y.,  Jin C.,  Fan J.,  Zhao X.,  Yu L.,   Du B.,  2016. pp 1667--1667,
  \mn@doi{10.1109/PIERS.2016.7734752}

\bibitem[\protect\citeauthoryear{{Xu}, {Wang}  \& {Chen}}{{Xu}
  et~al.}{2015}]{2015ApJ...798...40X}
{Xu} Y.,  {Wang} X.,   {Chen} X.,  2015, \mn@doi [\apj]
  {10.1088/0004-637X/798/1/40}, \href
  {http://adsabs.harvard.edu/abs/2015ApJ...798...40X} {798, 40}

\bibitem[\protect\citeauthoryear{{Yohana}, {Li}  \& {Ma}}{{Yohana}
  et~al.}{2019}]{2019arXiv190803024Y}
{Yohana} E.,  {Li} Y.-C.,   {Ma} Y.-Z.,  2019, {Yohana} E.,  {Li} Y.-C.,   {Ma}
  Y.-Z., , \href {https://ui.adsabs.harvard.edu/abs/2019arXiv190803024Y} {p.
  arXiv:1908.03024}

\bibitem[\protect\citeauthoryear{{Young}}{{Young}}{2002}]{2002AJ....124..788Y}
{Young} L.~M.,  2002, \mn@doi [\aj] {10.1086/341648}, \href
  {http://adsabs.harvard.edu/abs/2002AJ....124..788Y} {124, 788}

\bibitem[\protect\citeauthoryear{{Zhang}, {Bunn}, {Karakci}, {Korotkov},
  {Sutter}, {Timbie}, {Tucker}  \& {Wandelt}}{{Zhang}
  et~al.}{2016}]{2016ApJS..222....3Z}
{Zhang} L.,  {Bunn} E.~F.,  {Karakci} A.,  {Korotkov} A.,  {Sutter} P.~M.,
  {Timbie} P.~T.,  {Tucker} G.~S.,   {Wandelt} B.~D.,  2016, \mn@doi [\apjs]
  {10.3847/0067-0049/222/1/3}, \href
  {http://adsabs.harvard.edu/abs/2016ApJS..222....3Z} {222, 3}

\bibitem[\protect\citeauthoryear{{Zheng} et~al.,}{{Zheng}
  et~al.}{2017}]{2017MNRAS.464.3486Z}
{Zheng} H.,  et~al., 2017, \mn@doi [\mnras] {10.1093/mnras/stw2525}, \href
  {https://ui.adsabs.harvard.edu/abs/2017MNRAS.464.3486Z} {464, 3486}

\bibitem[\protect\citeauthoryear{{Zhu}, {Pen}, {Yu}  \& {Chen}}{{Zhu}
  et~al.}{2018}]{2018PhRvD..98d3511Z}
{Zhu} H.-M.,  {Pen} U.-L.,  {Yu} Y.,   {Chen} X.,  2018, \mn@doi [\prd]
  {10.1103/PhysRevD.98.043511}, \href
  {https://ui.adsabs.harvard.edu/abs/2018PhRvD..98d3511Z} {98, 043511}

\bibitem[\protect\citeauthoryear{{Zwaan} et~al.,}{{Zwaan}
  et~al.}{2004}]{2004MNRAS.350.1210Z}
{Zwaan} M.~A.,  et~al., 2004, \mn@doi [\mnras]
  {10.1111/j.1365-2966.2004.07782.x}, \href
  {http://adsabs.harvard.edu/abs/2004MNRAS.350.1210Z} {350, 1210}

\bibitem[\protect\citeauthoryear{{Zwaan}, {Meyer}, {Staveley-Smith}  \&
  {Webster}}{{Zwaan} et~al.}{2005}]{2005MNRAS.359L..30Z}
{Zwaan} M.~A.,  {Meyer} M.~J.,  {Staveley-Smith} L.,   {Webster} R.~L.,  2005,
  \mn@doi [\mnras] {10.1111/j.1745-3933.2005.00029.x}, \href
  {http://adsabs.harvard.edu/abs/2005MNRAS.359L..30Z} {359, L30}

\bibitem[\protect\citeauthoryear{{de Oliveira-Costa}, {Tegmark}, {Gaensler},
  {Jonas}, {Landecker}  \& {Reich}}{{de Oliveira-Costa}
  et~al.}{2008}]{2008MNRAS.388..247D}
{de Oliveira-Costa} A.,  {Tegmark} M.,  {Gaensler} B.~M.,  {Jonas} J.,
  {Landecker} T.~L.,   {Reich} P.,  2008, \mn@doi [\mnras]
  {10.1111/j.1365-2966.2008.13376.x}, \href
  {https://ui.adsabs.harvard.edu/abs/2008MNRAS.388..247D} {388, 247}

\makeatother
\end{thebibliography}

\end{document}